\documentclass[11pt]{article}
\usepackage{graphicx} % Required for inserting images
\usepackage[a4paper, margin = 1.0in]{geometry}
\usepackage[section]{placeins}
\usepackage{amsfonts}
\usepackage{amsmath}
\usepackage{float}
\usepackage[round]{natbib}
\usepackage{authblk}
\usepackage{url}
\usepackage{parskip}
\usepackage{booktabs}
\usepackage{float}
\usepackage{caption}
\usepackage{subcaption}
\floatstyle{plaintop}
\restylefloat{table}
\usepackage{threeparttable}
\usepackage{xcolor}
\usepackage{hyperref}
\definecolor{cb-green-sea}  {RGB}{  0, 146, 146}
\definecolor{cb-burgundy}   {RGB}{146,   0,   0}
\definecolor{cb-green-lime} {RGB}{ 36, 255,  36}
\usepackage{amssymb}% http://ctan.org/pkg/amssymb
\usepackage{pifont}% http://ctan.org/pkg/pifont
\newcommand{\cmark}{\color{cb-green-sea}\ding{51}}%
\newcommand{\xmark}{\color{cb-burgundy}\ding{55}}%

\newcommand{\mycomment}[1]{}
\title{DGP-LVM: Derivative Gaussian process latent variable models} 
\author[1,2,3]{Soham Mukherjee}
\author[2,3]{Manfred Claassen}
\author[1]{Paul-Christian Bürkner}
\affil[1]{Department of Statistics, Technical University of Dortmund, Dortmund, Germany}
\affil[2]{Department of Computer Science, University of Tübingen, T\"ubingen, Germany}
\affil[3]{University Hospital Tübingen, Faculty of Medicine, University of Tübingen, T\"ubingen, Germany}
\date{}
\begin{document}

\maketitle

\begin{abstract}
    We develop a framework for derivative Gaussian process latent variable models (DGP-LVMs) that can handle multi-dimensional output data using modified derivative covariance functions. The modifications account for complexities in the underlying data generating process such as scaled derivatives, varying information across multiple output dimensions as well as interactions between outputs. Further, our framework provides uncertainty estimates for each latent variable samples using Bayesian inference. Through extensive simulations, we demonstrate that latent variable estimation accuracy can be drastically increased by including derivative information due to our proposed covariance function modifications. The developments are motivated by a concrete biological research problem involving the estimation of the unobserved cellular ordering from single-cell RNA (scRNA) sequencing data for gene expression and its corresponding derivative information known as RNA velocity. Since the RNA velocity is only an estimate of the exact derivative information, the derivative covariance functions need to account for potential scale differences. In a real-world case study, we illustrate the application of DGP-LVMs to such scRNA sequencing data. While motivated by this biological problem, our framework is generally applicable to all kinds of latent variable estimation problems involving derivative information irrespective of the field of study. 
\end{abstract}

\section{Introduction} \label{sec-intro}

Gaussian processes (GPs) are a class of statistical models known for their flexible structure and favourable properties to analyse complex data \citep{williams_gaussian_1995}. Since their inception, several extensions have been proposed, among which the most relevant ones to this paper are adding derivative information to GPs \citep{solak_derivative_2002, gp_rasmussen_williams_2006}, building GPs with multiple outputs \citep{Cressie1993, teh_semiparametric_2005, gp_rasmussen_williams_2006} as well as modelling latent input variables \citep{lawrence_gaussian_2003, lawrence_probabilistic_2005}. Since differentiation is a linear operation, any variable and its derivative would be linearly related. For the same reason, as a fundamental property of GPs, a derivative of a GP is just another GP with a related covariance function. Together, this results in a single GP model for the outputs and their derivatives with a joint derivative covariance function \citep{solak_derivative_2002}. 

Derivative GPs have usually been studied and designed to model a single vector-valued output and have not been extended for multiple outputs. Multi-output GPs are most suitable when the response or output of the model contains multiple features, each expressed by its own dimension. One could fit individual GPs for each feature but then risks substantial loss of information in case of interactions between features. Thus a two-fold covariance structure was suggested in \citep{teh_semiparametric_2005} allowing GPs to account for this shared information between features. We extend this two-fold covariance structure to derivative GPs. As we will motivate more in a bit, our primary aim is to estimate latent (input) variables from observed input variables measured with error and output variables connected to the inputs via GPs; a challenge leading to what are called latent GPs \citep{lawrence_gaussian_2003}. When using derivative GPs, such latent inputs are shared between the original outputs and their derivative counterparts, effectively doubling the amount of information available for estimating the latent inputs. 

In real-world data, the derivatives are seldom exactly computed, which adds a major challenge to the modelling endeavour. If derivatives are computed with respect to the observed (non-latent) inputs, they are naturally only an approximation of the derivatives with respect to latent inputs. Conversely, if the derivatives are (implicitly) computed with respect to the latent inputs, the uncertainty of the latter will induce hard-to-quantify uncertainty in the estimated derivatives. One way or another, this lack of exact derivative information poses a serious challenge for derivative GP modelling. Existing approaches are not equipped to deal with the significant scale differences between outputs and derivatives, thus requiring modifications in its covariance functions to ensure valid and efficient latent variable estimation.

In this paper, we demonstrate a combination of all the above model extensions leading to our DGP-LVM: derivative Gaussian process latent variable model framework. We provide more context about the real-world modelling challenges in Section \ref{sec-motiv} as a basis for our motivation to develop DGP-LVM. The remainder of the paper is structured as follows. We discuss and provide context on related works in Section \ref{sec-rel-works}. We introduce our methodology and model development in Section \ref{sec-methods} and perform extensive simulation studies in Section \ref{sec-Sim} that demonstrate the relevance of our contributions. We further illustrate DGP-LVM on a real single-cell RNA sequencing data in Section \ref{sec-case-study} before discussing our methods' limitations and future work in Section \ref{sec-discuss}.

\subsection{Motivation} \label{sec-motiv}

In developmental biology, to describe temporal biological processes, researchers use stochastic approaches to understand cellular progression, that is, how cells develop and undergo changes in their state throughout various stages over the period of time \citep{maamar_noise_2007, losick_stochasticity_2008, raj_nature_2008}. Currently, this problem is frequently tackled with single-cell RNA sequencing (scRNA-seq) technologies by analysing messenger-RNA (mRNA) molecule counts as a measure of gene expression \citep{haque_practical_2017}. The measured gene expression (also called expression levels) provide the necessary information about the nature of cells at a specific point of time, also known as cell states, as well as their changes over time. However, due to the experimental limitations of the current sequencing methods each cell gets destroyed in the measurement process and can therefore be observed only once. This situation makes it difficult to infer cell state transitions and the overall sequence of cell states of a temporal biological process. To that end, pseudotime ordering is a popular approach to describe such a biological process as a sequence of cell states along a time sequence \citep{trapnell_dynamics_2014}.

Single-cell gene expression data provides information about cell state snapshots. While conventional pseudotime ordering approaches operate only on cell state snapshots to estimate pseudotime, only recently, directional information about cell state changes (i.e., derivative information) has been accounted for in this task \citep{gupta_simulation-based_2022}. Here, we hypothesise and demonstrate empirically later that including directional information on cell state transitions increases the precision in estimating pseudotime. This directional information is available through a quantity known as RNA velocity that is estimated from the difference in unspliced and spliced gene expression levels over latent experimental time (not to be confused with pseudotime) \citep{la_manno_rna_2018, Bergen2020}. By construction, this RNA velocity estimates the derivative of spliced gene expression data with respect to time. Concretely, our aim is to enable using the combination of RNA gene expression and RNA velocity in a single probabilistic framework for pseudotime estimation. This combination of RNA gene expression and its corresponding RNA velocity requires a novel statistical model approach.

In order to model such data, certain requirements must be satisfied. Starting from support for multi-dimensional outputs that allows inclusion of several genes for each cell, the model should account for varying gene-specific information as well as possible biologically induced interactions among genes. Moreover, since RNA velocity is only a derivative \textit{estimation} of gene expression levels, they are frequently on a significantly different scale than the gene expression levels. Dealing with this scale difference is a challenge that the model must address in order to provide reliable pseudotime estimates. In this paper, we demonstrate that DGP-LVM is able to tackle all of the above challenges and can estimate latent input variables with significantly higher accuracy than other GP models. Thus, we also demonstrate its potential to be applicable to estimating pseudotime through RNA gene expressions and their corresponding RNA velocities.

\subsection{Contributions} \label{sec-contrib}

\begin{itemize}
    \item We develop a probabilistic GP modelling framework for latent (input) variable estimation using derivative information for any multi-dimensional data-generating process. Our model accounts for dimension specific information and interactions between dimensions in a multi-dimensional data scenario which are common in (but not limited to) fields like single-cell biology.
    \item We develop a custom derivative structure for Squared Exponential (SE) and Matern class of covariance functions that is able to account for significant scale differences between the outputs and its corresponding derivatives. 
    \item Through extensive simulations, we demonstrate that our model provides substantially more accurate latent variable estimates than other GP models under realistic scenarios. 
    \item We showcase the application of our modelling approach on a reduced real-world scRNA-seq data set.
\end{itemize}

\section{Related work} \label{sec-rel-works}

Gaussian processes, as a class of models, underwent a wide range of extensions over the years giving rise to various forms of GP models. Specifically, three broad extensions relevant to this work are GPs with derivative information, multi-output GPs and GPs for latent variable modelling. Using derivative information for Gaussian processes was introduced in \cite{solak_derivative_2002} who replaced standard covariance functions with their derivative counterparts. This paved the way to modelling data along with its derivatives as a single GP model. Recently, derivative GPs were extended to support multiple inputs and scalable approximations \citep{eriksson_scaling_2018, padidar_scaling_2021}. In case of multi-output GPs, recent works \citep{moreno-munoz_heterogeneous_2018, joukov_fast_2022} study GPs with support for multiple outputs that are of varying nature in terms of data types, however multi-output GPs with derivative information have not been studied in detail yet. Latent GPs were introduced by \cite{lawrence_gaussian_2003, lawrence_probabilistic_2005} and have, so far, been predominantly used for dimensionality reduction \citep{titsias_variational_2009, titsias_bayesian_2010}. More recently, for similar applications of dimension-reduction technique, extensions on GP latent variable model for non-Gaussian likelihoods with different types of latent input structures were discussed in \cite{lalchand_generalised_2022}. These works also focus on scalable approximations to latent GPs. In contrast, we focus on estimating latent inputs that probabilistically explains a dependent multi-output variable.

For the modelling of scRNA-seq data, GPs have been broadly applied in two relevant directions, specifically, for clustering \citep{buettner_novel_2012,buettner_computational_2015} and temporal modelling \citep{hensman_hierarchical_2013}. Considering the latter, pseudotime estimation constitutes a major research direction as it is directly related to understanding the true underlying biological processes. It has been shown previously that point estimates of pseudotime are highly prone to infer false cellular ordering, thus suggesting Bayesian inference to provide uncertainty estimates alongside each estimated pseudotime \citep{campbell_bayesian_2015}. Further works focus on latent pseudotime estimations \citep{pseudotime_estimation_reid_john_wernisch, campbell_descriptive_2018} along with branching structures for trajectory inference \citep{ahmed_grandprix_2019} based on a GP framework. One of the main limitations in these works lie in their restricted use of gene expressions, taking into account expression profile snapshots as the only available information regarding cellular ordering. We provide evidence that including RNA velocity as derivative information holds the power to estimate latent pseudotime with increased precision compared to what previous approaches could achieve.

\section{Methods} \label{sec-methods}

We develop DGP-LVM, a framework for derivative Gaussian process modelling with the primary goal of estimating latent variables serving as (implicit) inputs to the GP. As the general setup, we consider a pair of variables $(y, x)$ where $y$ is the output (response) variable and $x$ is the input variable (covariate), with individual observations denoted as $y_i, x_i \in \mathbb{R}, i = \{1,...,N\}$ where $N$ is the number of observations. In addition to $y$ itself, we incorporate the derivative outputs $y' = \delta y / \delta x$ into the model. The components of DGP-LVM are first discussed individually, before we combine them into a single model.

\subsection{Derivative Gaussian processes} 
\label{sec-Deriv-GP}

A GP is a stochastic process specified by a mean function $m = m(x)$, and a covariance function $K = K(x, x^T)$, where $x^T$ indicate transpose of $x$, such that a finite set of these points will follow a multivariate Gaussian distribution \citep{williams_gaussian_1995}. Concretely, we consider GPs $f(x)$ such that $f(x) \sim \mathcal{GP}(m, K)$. Here we consider a constant mean function (similar to an intercept in regression models). If the output variable $y$ is univariate, modelling the relationship of $x$ and $y$ via a (single-output) GP and independent additive noise can be written as
\begin{equation} \label{gp model 1}
    y_i  = f(x_i) + \varepsilon_i,
\end{equation}
where $\varepsilon_i$ is the $i^{th}$ sample of $\varepsilon \sim \mathcal{N}(0, \sigma^2)$ assuming equal-variance Gaussian noise.  Together, this is equivalent to 
\begin{equation} \label{gp model 2}
    y_i \sim \mathcal{N}(f(x_i), \sigma^2).
\end{equation}
For, $i \neq j$ we have $\text{Cov}(y_i, y_j) = K(x_i, x_j)$ and for $i=j$, we have $\text{Cov}(y_i, y_j) = \text{Var}(y_i) = K(x_i, x_j) + \sigma^2$. The above notation will be extended to multi-output GPs in Section \ref{sec-multi-output}.

GPs are able to take advantage of derivatives in addition to its corresponding sample data to increase model accuracy. Since differentiation is a linear operator, a derivative of a GP is just another GP \citep{solak_derivative_2002, gp_rasmussen_williams_2006}. This property of GPs can be utilised to take derivative of a joint covariance structure of both $y$ and $y'$, if the second order derivative of the covariance function exists. The (joint) derivative GP is then given by
\begin{equation} \label{deriv GP}
    \left(\begin{matrix}
        f(x) \\
        f'(x)
    \end{matrix}\right)
    \sim
    \mathcal{GP}\left(
    \left(\begin{matrix}
        m_f \\
        m_{f'}
    \end{matrix}\right),
    %\left(\begin{matrix}
    %    0 \\
    %    0
    %\end{matrix}\right), 
    \left(\begin{matrix}
        K & K'\\
        K'^T & K''
    \end{matrix}\right)\right),
\end{equation}
where $m_f$ and $m_{f'}$ are constant mean functions corresponding to GP $f$ and it's derivative $f'$ respectively. $K'$ is the first derivative of the covariance function $K = K(x, x^T)$ with respect to $x$ and $K'^T$ is the first derivative of the covariance function with respect to $x^T$. $K''$ is the second order partial derivative of $K$ differentiating both with respect to $x$ and $x^T$. In other words, differentiation simply propagates through the covariance function (see Appendix A for mathematical details). The specific properties of such derivative GP models depend on the chosen covariance function. A common choice is the Squared Exponential (SE) covariance function with hyperparameters $\rho$ as length scale and $\alpha$ as the GP marginal standard deviation (SD). The derivative version of the SE covariance function is given by
\begin{equation} \label{k}
    K(x_i, x_j) = \alpha^2 \exp \left(-\frac{(x_i - x_j)^2}{2\rho^2}\right),
\end{equation}
\begin{equation} \label{k'}
    K'(x_i, x_j)= \alpha^2 \frac{(x_i - x_j)}{\rho^2} \exp \left(-\frac{(x_i - x_j)^2}{2\rho^2}\right),
\end{equation}
\begin{equation} \label{k''}
    K''(x_i, x_j) = \frac{\alpha^2}{\rho^4}(\rho^2 - (x_i - x_j)^2) \exp\left(-\frac{(x_i - x_j)^2}{2\rho^2}\right).
\end{equation}
Derivative covariance functions are obtainable generally for any chosen covariance function whose second order derivative exists. In this paper, we focus on SE and Matern class covariance functions as perhaps the most common choices. We provide further details on the mathematical forms of derivative SE, Matern 3/2 and Matern 5/2 covariance functions in Appendix A.

\subsubsection{Customised hyperparameters}
\label{sec-custom hyp}

Properly including the derivative observations $y'$ requires more than just using a basic derivative covariance function. Due to the properties of differentiation, $y'$ can be on a fundamentally different scale than $y$ and thus needs to be treated as such. In addition to having different signals (i.e., different GP components $f(x)$ vs. $f'(x)$), the error SDs for $y$ and $y'$ will also be different. Moreover, for real data, the observations $y'$ containing derivative information may not be the exact same as the true derivatives $ \delta y / \delta x$, but only be proportional to them (see Section \ref{sec-intro}). This proportionality induces a scale difference between $y'$ and what is canonically modelled by a basic derivative covariance function. This creates a major issue for models ignoring scale differences as we demonstrate in our simulations. 

To incorporate these scale considerations into our model, we propose to adjust the covariance function hyperparameters. Again using the SE covariance function as an example, we propose to introduce a second marginal SD parameter $\alpha'$, corresponding to the derivative part of the GP, while $\alpha$ now only concerns the original part of the GP:
\begin{equation} \label{custom k}
    K(x_i, x_j) = \alpha^2 \exp \left(-\frac{(x_i - x_j)^2}{2\rho^2}\right),
\end{equation}
\begin{equation} \label{custom k'}
    K'(x_i, x_j)= \alpha \alpha' \frac{(x_i - x_j)}{\rho^2} \exp \left(-\frac{(x_i - x_j)^2}{2\rho^2}\right),
\end{equation}
\begin{equation} \label{custom k''}
    K''(x_i, x_j) = \frac{\alpha'^2}{\rho^4}(\rho^2 - (x_i - x_j)^2) \exp\left(-\frac{(x_i - x_j)^2}{2\rho^2}\right).
\end{equation}
In other words, since $K(x_i, x_j) = Cov(y_i, y_j)$ we account for the GP marginal variance through $\alpha^2$ since we are only concerned with the original part of the GP $f$. In the case of $Cov(y_i, y'_j)$ and $Cov(y'_i, y_j)$, we compute $K'(x_i, x_j)$ and thus we account for the GP marginal variance through $\alpha\alpha'$ where $\alpha$ belongs to $f$ and $\alpha'$ belongs to $f'$. Finally, in the case of $Cov(y'_i, y'_j)$, we compute $K''(x_i, x_j)$ and are thus accounting for the GP marginal variance through $\alpha'^2$ since we are only dealing with the derivative part of the GP $f'$. Since, in addition to the product term $\alpha\alpha'$ in $K'$, we also estimate $\alpha^2$ and $\alpha'^2$ separately through $K$ and $K''$, respectively, estimating both $\alpha$ and $\alpha'$ does not cause identifiability issues.

Similarly, we define two residual standard deviation parameters $\sigma$ and $\sigma'$, accounting for measurement noise in $y$ and $y'$, respectively. The scale of $\rho$ is only dependent on the scale of $x$, which is constant across outputs and their derivatives, such that $\rho$ does not need to be split up into two parameters. Together, independent of specifically chosen covariance function, the DGP-LVM on $y$ and $y'$ with independent, additive Gaussian noise is then specified as
\begin{equation} \label{custom DGP-LVM model}
y_i \sim \mathcal{N}(f(x_i), \sigma^2) \quad \text{and} \quad y'_i \sim \mathcal{N}(f'(x_i), \sigma'^2).
\end{equation}

\subsection{Multidimensional outputs} 
\label{sec-multi-output}

Multivariate output GPs (or multi-output GPs) model multiple response variables $\{y_1,...,y_D\}$ jointly over $D>1$ output dimensions \citep{gp_rasmussen_williams_2006}. Extending our univariate notation, the individual output values are now denoted as $y_{di}$ for dimension $d$ and observation $i$, with corresponding derivative values $y'_{di}$. Multi-output GPs are created by first setting up $D$ independent, univariate Gaussian processes $f_d(x)$ each with their own set of hyperparameters, that is, $(\rho_d, \alpha_d$, $\alpha'_d$, $\sigma_d$ and $\sigma'_d)$ for Matern class of covariance functions with adjusted scales. Subsequently the univariate GPs are related to one another by folding them with a ($D$-dimensional) across-dimension correlation matrix $C$ \citep{teh_semiparametric_2005, bonilla_multi-task_2007}. That is, for each observation $i$, we obtain a vector of across-dimension correlated GP values as
\begin{equation}
\label{multi-gps}
\left(\begin{array}{c} \tilde{f}_1(x_i) \\ \dots \\ \tilde{f}_D(x_i) \end{array}\right) 
= L \times \left(\begin{array}{c} f_1(x_i) \\ \dots \\ f_D(x_i) \end{array}\right),
\end{equation}
where $L$ is the Cholesky factor of $C$, that is, $C = LL^T$ with $L$ being lower-triangular. This way, multi-output GPs combine two dependency structures, one within dimensions (and across observations) as expressed by the univariate GPs through corresponding covariance functions and one across output dimensions (but within observations) as expressed by $C$ (or $L$).

This readily generalizes to derivative GPs by applying Equation~\eqref{multi-gps} to the derivative GP values $f'_d(x_i)$ as well, which results in the across-dimension correlated values $\tilde{f}'_d(x_i)$. Adding independent Gaussian noise, our derivative multi-output GP model then implies for all $d$ and $i$:
\begin{equation} \label{multi-derivGPs}
y_{di} \sim \mathcal{N}(\tilde{f}_d(x_i), \sigma_d^2) \quad \text{and} \quad 
y'_{di} \sim \mathcal{N}(\tilde{f}'_d(x_i), \sigma_d'^2).
\end{equation}
 
\subsection{Latent variable inputs} \label{sec-latent-inputs}

So far, we have considered the input $x$ to be known exactly. However, in practice, we often only have a noisy measurement $\tilde{x}$ of $x$ available. In this context, the true $x$ becomes a latent variable, which needs to be appropriately modelled and subsequently estimated. If we assume that the measurements $\tilde{x}$ are Gaussian with known measurement SD $s$, we can write for each observation $i$:
\begin{equation} \label{eqn-latent prior}
    \tilde{x}_i \sim \mathcal{N}(x_i, s^2).
\end{equation}
The implied latent $x_i$ is then passed to the GP covariance function, which results in what is known as latent(-input) GPs \citep{lawrence_gaussian_2003, lawrence_probabilistic_2005, titsias_bayesian_2010}. Such latent-input GPs are even harder to fit than their non-latent counterparts: Not only does the number of unknowns increase substantially, but also new identification issues arise due to both $x$ and $\rho$ now being unknown (see Section \ref{sec-full-model} for details on how we deal with this). 

\subsection{The full model} 
\label{sec-full-model}

Below, we summarize all the extensions that together make up our proposed DGP-LVM framework. To shorten the notation, let us denote the vector of GP hyperparameters for dimension $d$ as $\theta_d$. For our considered covariance functions, $\theta_d$ includes the length scale $\rho_d$, GP marginal SDs $\alpha_d$ and $\alpha'_d$ as well as the error SDs $\sigma_d$ and $\sigma'_D$. Considering the SE covariance function as an example, full DGP-LVMs are then specified as follows:
\begin{equation} \label{full-dgplvm}
\begin{aligned}
  \left(\begin{matrix}
        f_d(x) \\
        f'_d(x)
    \end{matrix}\right)
    &\sim
    \mathcal{GP}\left(
    \left(\begin{matrix}
        m_f \\
        m_{f'}
    \end{matrix}\right),
    \left(\begin{matrix}
        K_d & K_d'\\
        K_d'^T & K_d''
    \end{matrix}\right)\right), \\[3pt]
    K_d(x_i, x_j) &= \alpha_d^2 \exp \left(-\frac{(x_i - x_j)^2}{2\rho_d^2}\right), \\[3pt]
    K'_d(x_i, x_j) &= \alpha_d \alpha_d' \frac{(x_i - x_j)}{\rho_d^2} \exp \left(-\frac{(x_i - x_j)^2}{2\rho_d^2}\right), \\[3pt]
    K''_d(x_i, x_j) &= \frac{\alpha_d'^2}{\rho_d^4}(\rho_d^2 - (x_i - x_j)^2) \exp\left(-\frac{(x_i - x_j)^2}{2\rho_d^2}\right), \\[3pt]
    y_{di} &\sim \mathcal{N}(f_d(x_i), \sigma_d^2), \\[3pt]
    y'_{di} &\sim \mathcal{N}(f'_d(x_i), \sigma_d'^2), \\[3pt]
    \tilde{x}_i &\sim \mathcal{N}(x_i, s^2), \\[3pt]
    \theta_d &\sim p(\theta_d) = p(\rho_d) \, p(\alpha_d) \, p(\alpha'_d) \, p(\sigma_d) \, p(\sigma'_d).    
\end{aligned}
\end{equation}
Following the above specifications, after marginalizing out $f$ and $f'$, the multi-output joint probability density factorizes as 
\begin{equation}
\label{eqn-jt prob model}
    p(y, y', x, \theta) = \prod_d^D p(y_d \mid x, \theta_d) \, p(y'_d \mid x, \theta_d) \, p(x) \, p(\theta_d).
\end{equation}
where $p(y_d \mid x, \theta_d)$ and $p(y'_d \mid x, \theta_d)$ denote the respective GP-based likelihoods for a single output dimension. $p(x)$ denotes the prior for the latent $x$ implied by the measurement model Eq.\eqref{eqn-latent prior}. More details on the choice of prior distributions are discussed in Section \ref{sec-model-setup}. Using Bayes' rule, we obtain the joint posterior over $x$ and $\theta$ as
\begin{equation}
\label{eqn-post x}
    p(x, \theta \mid y, y') = 
    \frac{p(y, y', x, \theta)}
    {\int\int p(y, y', x, \theta) \, dx \, d\theta}.
\end{equation}
Posterior samples of $x$ and $\theta$ (i.e. all the covariance function hyperparameters) are obtained through MCMC sampling via adaptive Hamiltonian Monte Carlo \citep{Neal2011HMC, HoffmanM2014NUTS}. We implemented all models in Stan using the RStan interface \citep{Stan_guide_2024}.

Implemented as above, DGP-LVMs can be applied to the aforementioned problem of pseudotime estimation from single-cell RNA sequencing data. The scRNA-seq data we consider consists of spliced RNA gene counts and RNA velocity, the time derivative of gene counts. DGP-LVM allows inclusion of both these information into a single model (see Section \ref{sec-Deriv-GP}). Since the RNA velocity is not an exact derivative of spliced RNA counts, it induces a scale difference that is solved by DGP-LVM as shown in Section \ref{sec-custom hyp}. Given that single-cell RNA sequencing data is multi-dimensional, DGP-LVM is designed as a multi-output model (see Section \ref{sec-multi-output}). The primary aim of DGP-LVM is to estimate the latent inputs as explained in Section \ref{sec-latent-inputs}, which perfectly aligns with pseudotime estimation since pseudotime is an unobserved cell ordering, and is hence considered as a latent variable. Moreover, the RNA sequencing data comes with its own cell capture time or experimental time which can be considered as a noisy version of the true pseudotime.

\section{Simulation study} \label{sec-Sim}

The fundamental issue for validating and comparing models designed to estimate latent variables is the lack of ground truth values for real-world data. Thus, it is crucial to test any latent variable model through extensive simulations where the ground truth is available and controllable. Below, we discuss and provide evidence for the importance of our proposed model innovations. Concretely, we showcase DGP-LVM on multiple simulated data setups that closely represent the complexities of a real scRNA sequencing data. 

\subsection{Simulated data} \label{sec-sim-data}

We consider five primary scenarios to generate simulated data. In our first scenario, we generate data from a multi-output GP with scaled derivative SE covariance function. In our second and third scenarios, we generate data from a multi-output GP but with scaled derivative Matern 3/2 and 5/2 covariance functions, respectively. All of the above scenarios constitute cases where the estimated DGP-LVMs align with the true underlying process. However, datasets generated this way can vary strongly in the amount of signal they contain, thus adding a lot of random variation in the simulation results. To account for this issue, in our fourth scenario, we generate data from a derivative periodic process with the true generating function
\begin{equation} \label{eqn:per-sim-f}
\begin{aligned}
    f_{ij} &= \alpha_j \sin \left(\frac{x_i}{\rho_j}\right) \\[3pt]
    f'_{ij} &= \frac{\alpha'_j}{\rho_j} \cos \left(\frac{x_i}{\rho_j}\right)
\end{aligned}
\end{equation}   
and corresponding data simulations as
\begin{equation} \label{eqn:per-sim-y}
    y_{ij} \sim \mathcal{N}(f_{ij}, \sigma^2) \quad \text{and} \quad y'_{ij} \sim \mathcal{N}(f'_{ij}, \sigma'^2).
\end{equation}
In all of these scenarios, the data is generated with varying hyperparameters and correlated outputs assumptions in play (see Section \ref{sec-methods}). The hyperparameters of the periodic data generating process fulfil a similar purpose to those of SE, Matern 3/2 and Matern 5/2 derivative GPs. Hence we choose to use the same hyperparameter names for simplicity. The scenario of periodic data (Eq.\eqref{eqn:per-sim-f}) is important on two counts. First, it allows us to better control the amount of signal contained in each generated dataset. Second, it demonstrates that DGP-LVM can achieve good results even when the underlying generating process in not actually a GP. 

Additionally, in the fifth scenario, we further increase the complexity by adding a quadratic and linear trend to the above periodic and corresponding derivative functions, respectively, as
\begin{equation} \label{eqn:per-trend-sim-f}
\begin{aligned}
    f_{ij} &= \alpha_j \sin \left(\frac{x_i}{\rho_j}\right) + bx_i^2, \\[3pt]
    f'_{ij} &= \frac{\alpha'_j}{\rho_j} \cos \left(\frac{x_i}{\rho_j}\right) + 2bx_i,
\end{aligned}
\end{equation} 
where $b$ is a scaler and is chosen in accordance with the acting periodicity parameter $\rho$. The data generating process then follows Eq. \eqref{eqn:per-sim-y}. This fifth scenario is designed to test the limitations of modelling non-stationary data with stationary GPs.

To demonstrate the adversity of scale difference between $y$ and $y'$, we induce a scaling factor of $\lambda = 3$ that propagates through the GP marginal SD and error SD (see Section \ref{sec-custom hyp}). The GP marginal SD for the output $y$ and the derivative $y'$ are related through $\lambda$ such that $\alpha = \lambda \alpha'$. Similarly for the error SD, $\sigma = \lambda \sigma'$. Therefore, we only specify sampling distributions for $\alpha' \sim \text{Normal}^+(3, 0.25^2)$ and $\sigma' \sim \text{Normal}^+(1, 0.25^2)$.  In reality, $\lambda > 3$ or $\lambda < 1/3$ may very well occur (see Section \ref{sec-case-study}). In the simulations, we chose to avoid more extreme $\lambda$ values to prevent substantial convergence issues for models without scaling modifications. This allows us to showcase these models' (reduced) performance without confounding this finding with convergence considerations. The ground truth GP length scale is sampled as $\rho \sim \text{Normal}^+(1, 0.05^2)$. The choice of our sampling distributions, especially for the length scale $\rho$ being an informed prior, enables us to explicitly select a range of values for our hyperparameters for which the simulated GP data would contain sufficient amount of signal. Further, we introduce an uniform between-dimension correlation of $0.5$ for all the simulated data scenario as to represent moderate interactions between outputs. Combined with the sampling of true hyperparameters for each output dimension, our simulated data mimics the real-world data scenario where such assumptions are prevalent.

Lastly, we generate ground truth for $x$ as a sequence of values between $\{0.5,..., 9.5\}$ with a total output sample size of $N = 20$ for $y$ and $y'$ each and we choose a value for prior measurement SD of the noisy $\tilde{x}$ as $s = 0.3$ (see Section \ref{sec-latent-inputs}). This resembles the realistic scenario where observed times are already a relatively good measure of latent pseudotime considering the overall input scale. Both simulation studies are generated as multi-output data with three sets of output dimensions namely $D = 2, 5 \text{ } \text{and} \text{ } 10$. To keep model estimation times within manageable bounds, each simulated dataset contains only 20 $y$ and correspondingly 20 $y'$ sample points. We perform 50 trials for each simulation scenario, that is, generate 50 datasets for the three GP data, the periodic and the periodic with trend scenario, respectively. 

\subsection{Model setup} \label{sec-model-setup}

The models within our DGP-LVM framework can vary specifically in four components, namely the inclusion of (1) derivative information, (2) scaled derivatives, (3) varying hyperparameters, as well as (4) correlations across output dimensions. In order to study the individual importance of the four components, we systematically enable/disable each of them and investigate the resulting models' performances. The underlying data generating process contains all of the above components, so any model that only has a subset of components will be misspecified at least to some degree. In our simulations, component combinations were fully crossed where sensible (see Table \ref{tab:model-comp} for an overview). We fit 12 GP models for each selected number of output dimensions $D$ resulting in a total of 36 models fitted per generated dataset for a specific simulation scenario. We only exclude specific, non-sensible combinations. For example, it does not make sense to ask if a GP model, which does not include derivative information, accounts for the scaling of the derivatives. We use a constant mean function (similar to an intercept in regression models) in all our models for both $f$ and $f'$(wherever applicable) throughout the simulation study. Such specifications help with overall location shifts in the data.

\begin{table}[!ht]
    \centering
    \begin{threeparttable}
    \begin{tabular}{cccc}
    \toprule
Derivative information & Scaled derivatives & Varying hyperparameters & Correlated outputs\\
 \midrule
\cmark & \cmark & \cmark & \cmark\\ 
\cmark & \cmark & \cmark & \xmark\\ 
\cmark & \cmark & \xmark & \cmark\\ 
\cmark & \xmark & \cmark & \cmark\\ 
\cmark & \cmark & \xmark & \xmark\\ 
\cmark & \xmark & \xmark & \cmark\\ 
\cmark & \xmark & \cmark & \xmark\\ 
\cmark & \xmark & \xmark & \xmark\\ 
\xmark & \xmark & \cmark & \cmark\\ 
\xmark & \xmark & \cmark & \xmark\\ 
\xmark & \xmark & \xmark & \cmark\\ 
\xmark & \xmark & \xmark & \xmark\\ 
 \bottomrule
    \end{tabular}
    \begin{tablenotes}
       \item \small \textit{Note: Each table row denotes the assigned modifications to the fitted models. The first row shows the modifications involved in DGP-LVM.}
    \end{tablenotes}
    \end{threeparttable}
    \caption{GP models along with their specifications used for simulated scenarios}
    \label{tab:model-comp}
\end{table} 

Prior specifications for all the model hyperparameters involved were aligned with the data generating conditions to a reasonable extent to better showcase the model contributions. We specify separate priors for the marginal SDs and error SDs corresponding to $f$, $f'$ and $y$, $y'$ respectively, to account for the scale differences between the original and derivative part of the data. We specify the priors for GP marginal SDs $\alpha \sim \text{Normal}^+(9, 0.75^2)$ and $\alpha' \sim \text{Normal}^+(3, 0.25^2)$. In case of error SDs we specify $\sigma \sim \text{Normal}^+(3, 0.75^2)$ and $\sigma' \sim \text{Normal}^+(1, 0.25^2)$. We shifted our priors for the original part of the GP $f$ and data $y$ in accordance with our choice of scale difference $\lambda$ in the simulation scenarios. We use an informative prior on the length scale $\rho \sim \text{Normal}^+(1, 0.05^2)$ is roughly based on the mean Euclidean distance between the $\tilde{x}_i$ (as in  Section \ref{sec-latent-inputs}) as well as the data generating specifications showed in Section \ref{sec-sim-data}. As prior for the between-output correlation matrix $C$, we apply an unimodal LKJ(1) distribution \citep{LKJcholesky2009} defined over the positive definite symmetric matrices with unit diagonals. This distribution is a common prior choice for correlation matrices. For our constant mean function (intercepts), we used a Normal distribution with data-specific Mean and SD (for both $y$ and $y'$ correspondingly) as prior.

\begin{figure}[!ht]
    \centering
    \includegraphics[scale=0.5]{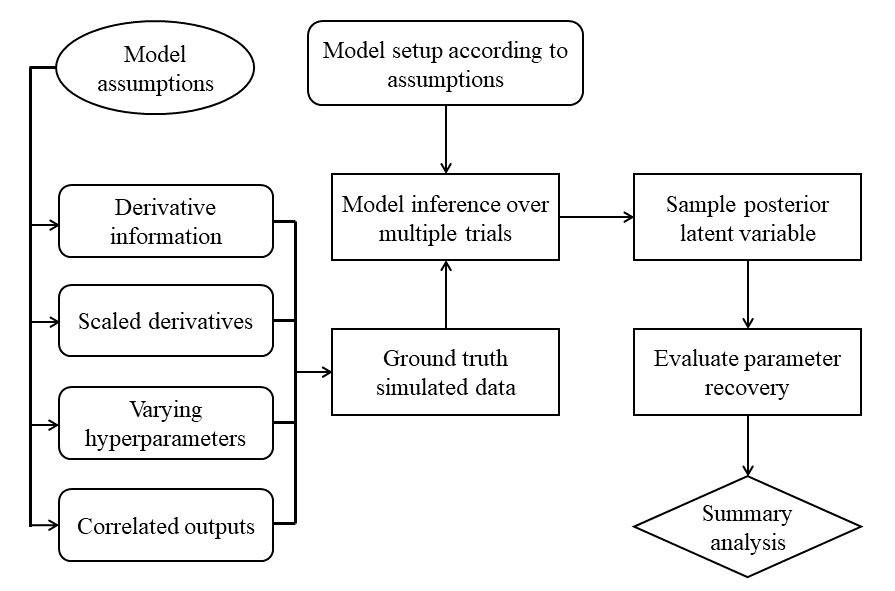}
    \caption{\textit{High-level overview of the simulation study design.}}
    \label{fig:sim-study-flowchart}
\end{figure}

\begin{figure}[!ht]
    \centering
    \includegraphics[width=\linewidth]{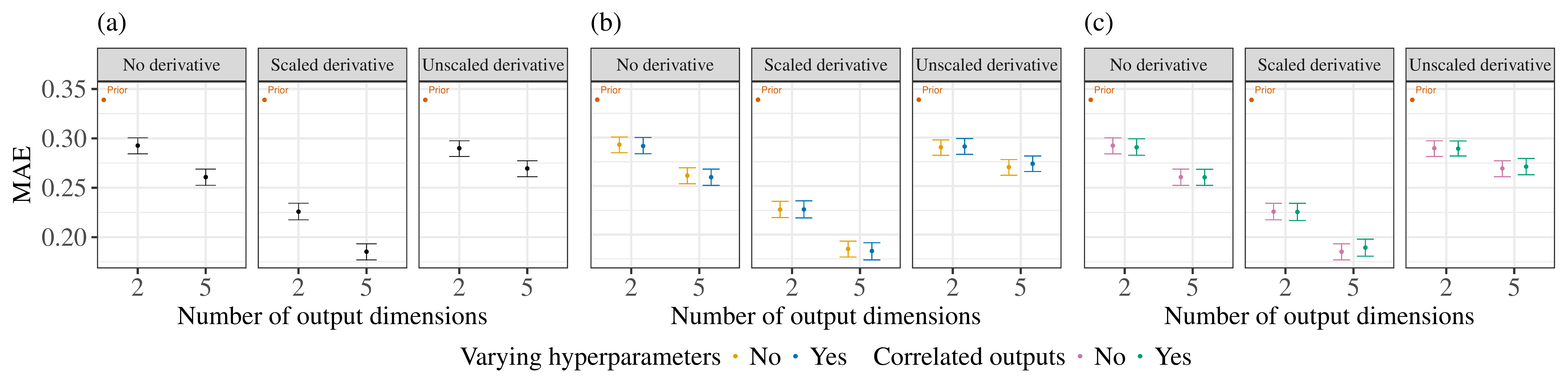}
    \caption{\textit{Squared exponential scenario with 4 MCMC chains: main effects of including (a) scaled derivatives and interaction effects of assuming (b) varying hyperparameters and (c) correlated outputs on recovery of latent inputs (full version) using MAE.}}
    \label{fig:app-se-chains-x-effects-mae-full} 
\end{figure}

All models were specified in Stan \citep{Stan_guide_2024} and fitted with a single MCMC chain of 3000 iterations in total with 1000 warm-ups. We decided to run only a single chain per model to reduce overall computation times. However, we show that using multiple chains yield similar results down the line (see Appendix: Figures \ref{fig:app-se-chains-x-effects-rmse-full} and \ref{fig:app-se-chains-x-effects-mae-full}.). All models were fitted on all generated datasets. The full study design is depicted in Figure \ref{fig:sim-study-flowchart}. 

The simulation studies were conducted using 50 vCPUs (Intel(R) Xeon(R) Gold 6230R CPU @ 2.10 GHz) with 720 GB memory allowances. The maximum runtime (in hours) for DGP-LVM with $D = 10$ (for a simulated dataset) were approximately 4.6 for the SE model, 5.9 for the Matern 3/2 model and 8.9 for the Matern 5/2 model.

\subsection{Summary methods}

In order to evaluate how well fitted models recover the latent ground truth, we compare posterior samples of the latent input variable $x$ denoted by $x_{post}$ with their respective ground truth values denoted by $x_{true}$ using the root mean squared error (RMSE):   
\begin{equation}\label{eqn-rmse}
    \text{RMSE}(x_{post}) = \sqrt{\mathbb{E}(x_{post} - x_{true})^2} 
\end{equation} 
and the mean absolute error:
\begin{equation}\label{eqn-mae}
    \text{MAE}(x_{post}) = \mathbb{E}(\mid x_{post} - x_{true}\mid)
\end{equation}
where the expectations are taken over the posterior (approximated via samples). In case of RMSE, $\mathbb{E}(x_{post} - x_{true})^2$ can be decomposed into $\text{Var}(x_{post})$ and $\text{Bias}(x_{post}, x_{true})^2$, thus measuring Bias-variance trade-off. We compute RMSE and MAE from all fitted models shown in Table \ref{tab:model-comp} for each set of output dimensions (2,5 and 10). We prefer models that provide both low bias indicating posterior mean estimates close to the ground truth as well as lower posterior variance indicating high precision, together resulting in an overall low RMSE. Similarly, we prefer low MAE since it shows the amount of absolute bias present while estimating the latent variable. Overall, RMSE penalizes the models for estimating outlying posteriors while MAE is more lenient in that sense. 

To analyse RMSE and MAE values, we use a multilevel analysis of variance model (ANOVA) fitted with brms \citep{burkner_brms_2017}, which disentangles the contributions of each model component. Using a multilevel model is important to account for the dependency between results of all models fitted on the same dataset. We model fixed main effects of scaled derivatives, varying hyperparameters, correlated outputs and number of output dimensions. For this purpose, we consider scaled derivatives as a factor variable with three levels corresponding to models that (a) do not include derivative information, (b) models that include derivative information with scaling and (c) models that include derivative information without scaling. Varying hyperparameters are represented by a binary factor variable that denotes varying vs. constant hyperparameters across output dimensions. Similarly, correlated outputs represented by a binary factor variable that indicates if the multiple outputs are assumed to be correlated or not. We additionally model fixed interaction effects between (a) scaled derivatives and varying hyperparameters and (b) scaled derivatives and correlated outputs. Since our simulation study is performed over 2, 5, and 10 output dimensions, we include dimension as factor variable with three levels and allowed it to interact with all previously mentioned (fixed) main and interaction effects. We account for the dependency structure in the RMSE and MAE values, induced by fitting multiple models to the same simulated dataset, by a random intercept over datasets as well as corresponding random slopes of the scaled derivatives, varying hyperparameters, and correlated outputs factors. Further, we account for the dependency in the evaluation metric values for the 20 latent inputs estimated from a single model through a random intercept per fitted model. The results based on RMSE are presented in Section \ref{sec-sim-results} while their corresponding MAE results are shown in Appendix B.

\subsection{Model convergence}

We investigate the convergence of our fitted GP models for all the five simulation scenarios mentioned before. To that end, we use standard MCMC sampling diagnostics including state-of-the-art versions of the scale reduction factor $\widehat{R}$, the bulk effective sample size (Bulk-ESS) and the tail effective sample size (Tail-ESS)  \citep{RankNorm_Vehtari_etal}. The combined check of these measures provide a comprehensive picture of individual parameter model convergence.

In general, $\widehat{R}$ should be very close to 1 and should ideally not exceed 1.01 \citep{RankNorm_Vehtari_etal}. In a simulation setup, we can evaluate the goodness of the posterior estimation also independently of convergence, as we have access to ground truth values. Hence, also in light of the relatively short MCMC chains, we decide to apply a more relaxed threshold of 1.1. Bulk-ESS indicates the reliability of measures of central tendency such as the posterior mean or median. Tail-ESS indicates the reliability of the 5\% and 95\% quantile estimates, which are commonly used to construct credible intervals. Both Bulk-ESS and Tail-ESS should have values greater than 100 times the number of MCMC chains. We computed all the convergence measures with the posterior package \citep{posterior2023}. 

\begin{figure}[!ht]
    \centering
    \includegraphics[width = \linewidth]{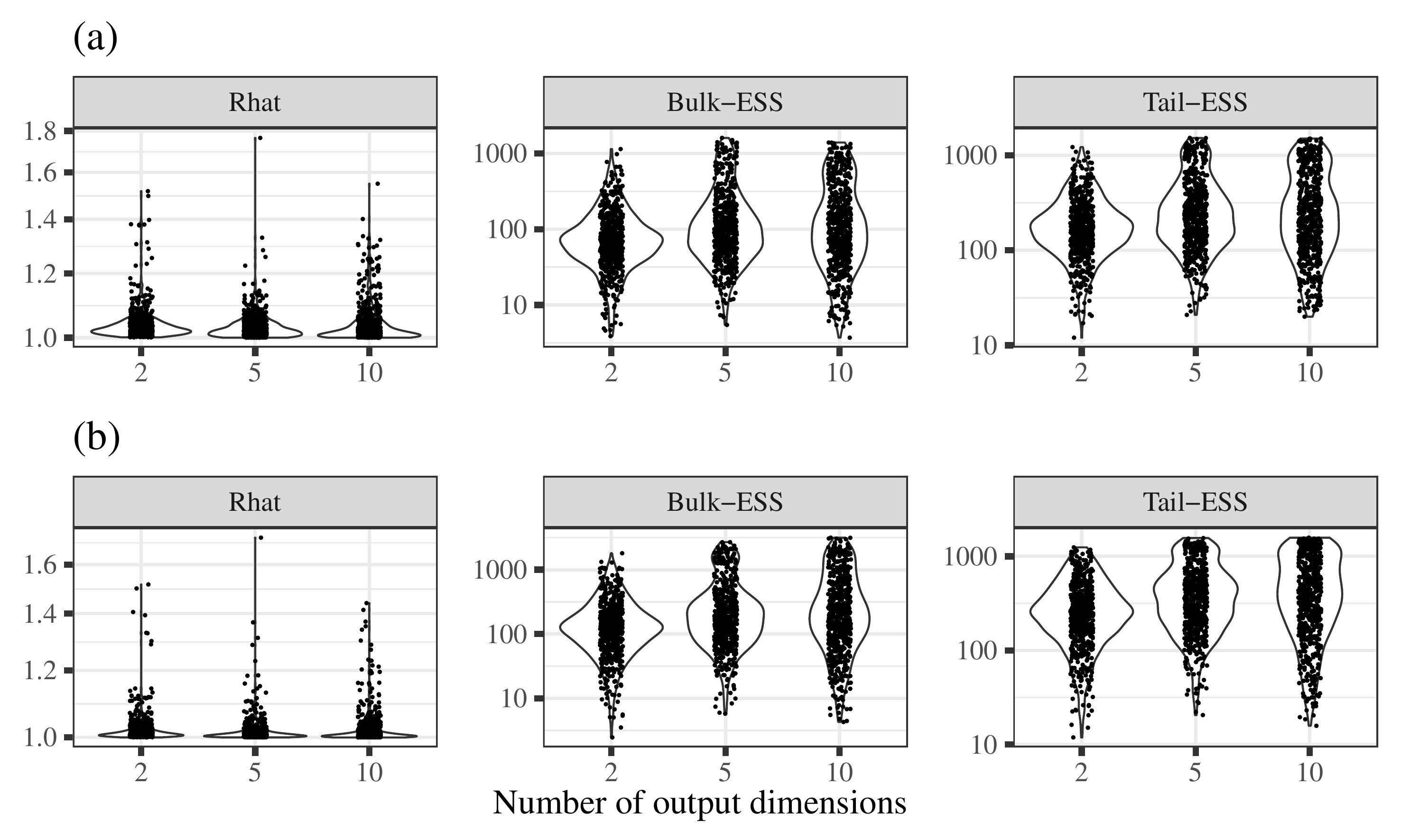}
    \caption{\textit{Squared exponential scenario: Convergence measures for (a) latent inputs and (b) GP hyperparameters. The individual points correspond to each fitted models per simulated data. The y-axis for Bulk and Tail ESS plots are log10 transformed.}}
    \label{fig:se-model-validation-gp}
\end{figure}

For latent inputs and hyperparameters obtained from the simulated SE data scenario, we show $\widehat{R}$, Bulk-ESS, and Tail-ESS in Figure \ref{fig:se-model-validation-gp}. We present the MCMC diagnostics plots for the other simulated data scenarios (see Figures \ref{fig:m32-model-validation-gp}-\ref{fig:per-trend-model-validation-gp}) in Appendix B owing to their similar nature. The $\widehat{R}$ were all satisfactory for majority of the simulation trials, with the exception of a few outlying models per trial. The GPs with derivative SE covariance function per simulation trial had better convergence as compared to the GPs with derivative Matern covariance functions. This was expected due to the increased model complexity of the Matern covariance functions. Moreover, the derivative Matern 3/2 (see Figure \ref{fig:m32-model-validation-gp}) being the boundary of existing derivative covariance functions among the Matern class makes it more complex for the sampler to perform as good as the Matern 5/2 (see Figure \ref{fig:m52-model-validation-gp}) and subsequently the much simpler SE covariance function. The Bulk-ESS and Tail-ESS were consistently higher than the suggested threshold for all the cases, thus satisfying the recommended criteria. The convergence results for the periodic and periodic with trend scenarios as shown in Figures \ref{fig:per-model-validation-gp} and \ref{fig:per-trend-model-validation-gp} were similar to derivative SE data simulation scenario.

\subsection{Results} \label{sec-sim-results}

For all of the simulation scenarios discussed in Section \ref{sec-sim-data}, we evaluated the effects of including derivative information, accounting for scale differences between $y$ and $y'$, estimating varying hyperparameters across multiple outputs as well as correlated outputs. We summarize these aforementioned conditions as model assumptions and show their effects on the RMSE as our primary model evaluation measure of the posterior estimates of latent $x$ and covariance function hyperparameters with respect to their true simulated values. The corresponding MAE results were qualitatively highly similar and are thus only shown in Appendix B.

\subsubsection{Model evaluation: Latent inputs}

In addition to the posterior model evaluation measure estimates obtained from multilevel ANOVA, we also show the prior RMSE that would be expected if we only used the prior measurement model $\tilde{x}_i \sim \mathcal{N}(x_i, s^2)$ to infer $x$. Consequently, the prior evaluation measure acts as a benchmark to illustrate how much precision we gain through the GP modelling of output data.

\begin{figure}[!ht]
    \centering
    \includegraphics[width = \linewidth]{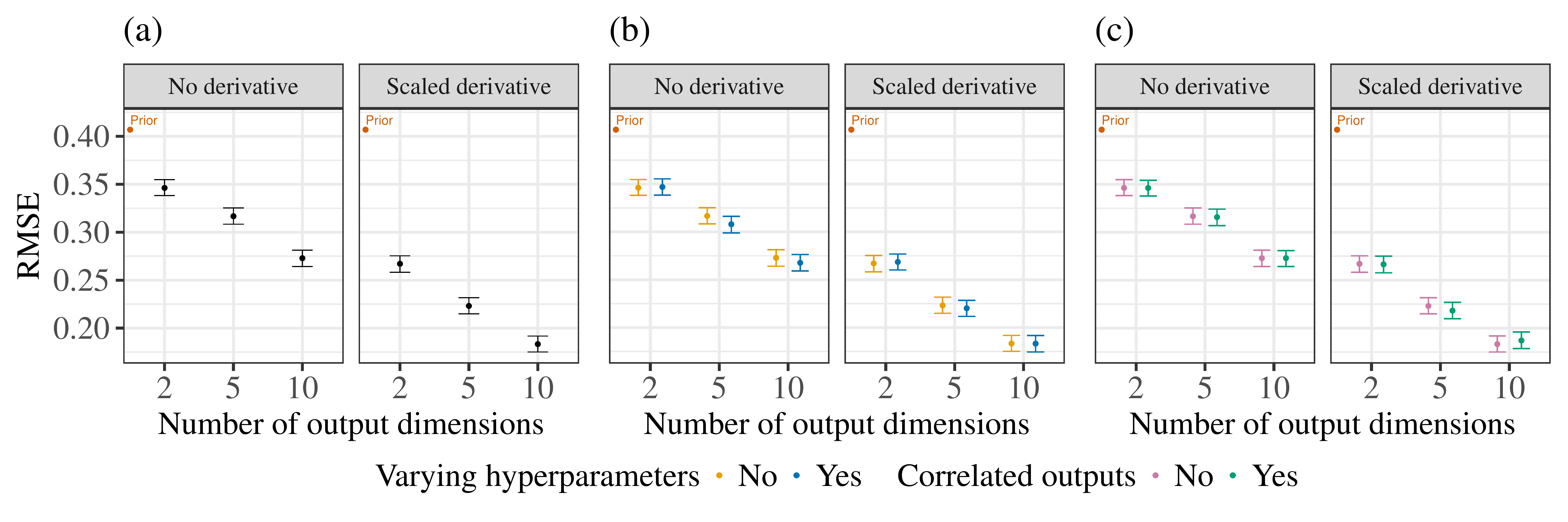}
    \caption{\textit{Squared exponential scenario: Main effects of including (a) scaled derivatives and interaction effects of assuming (b) varying hyperparameters and (c) correlated outputs on recovery of latent inputs}}
    \label{fig:se-sim-x-effects}
\end{figure}

\begin{figure}[!ht]
    \centering
    \includegraphics[width = \linewidth]{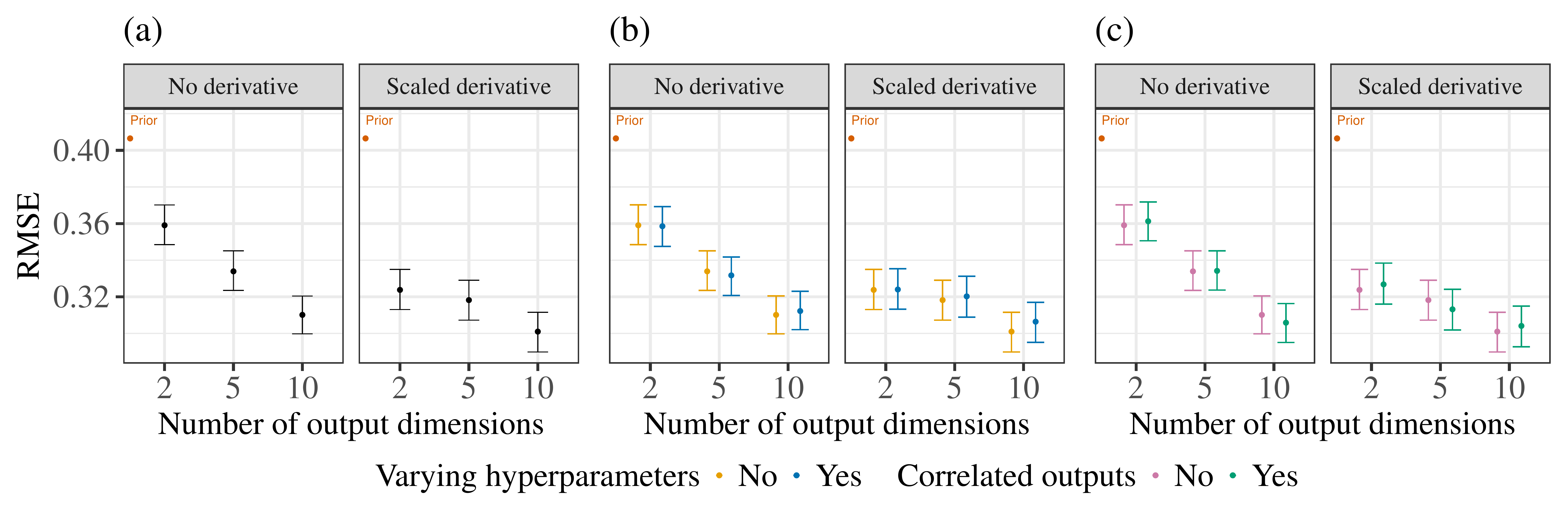}
    \caption{\textit{Matern 3/2 scenario: Main effects of including (a) scaled derivatives and interaction effects of assuming (b) varying hyperparameters and (c) correlated outputs on recovery of latent inputs}}
    \label{fig:m32-sim-x-effects}
\end{figure}

\begin{figure}[!ht]
    \centering
    \includegraphics[width = \linewidth]{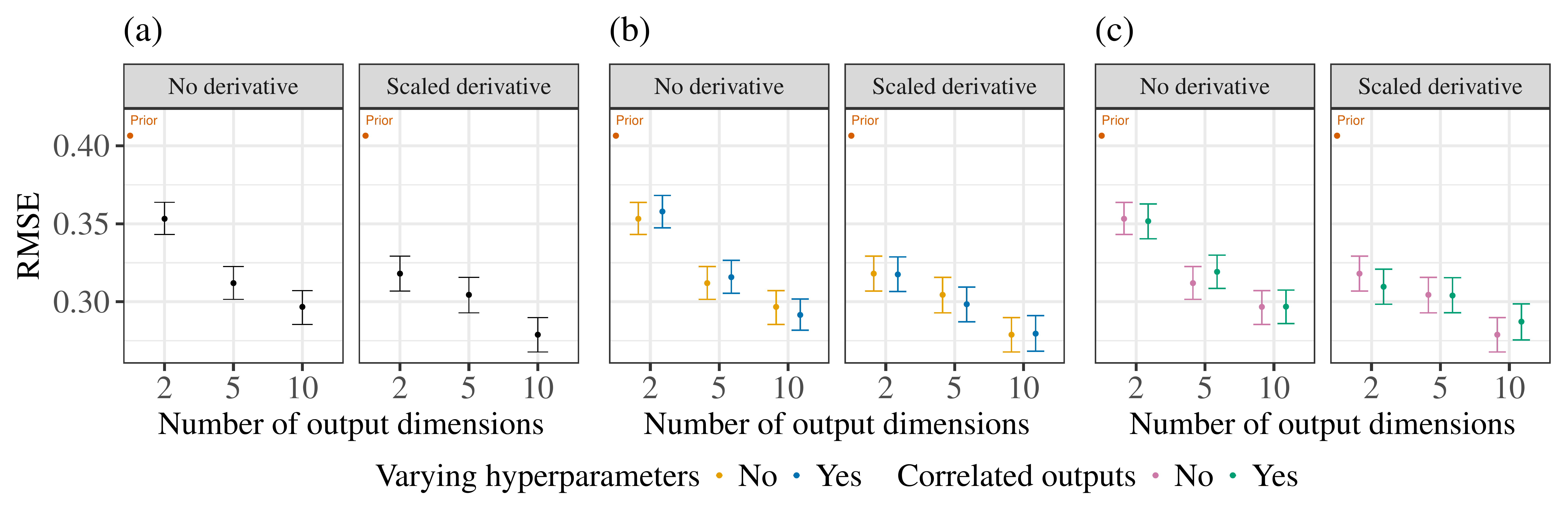}
    \caption{\textit{Matern 5/2 scenario: Main effects of including (a) scaled derivatives and interaction effects of assuming (b) varying hyperparameters and (c) correlated outputs on recovery of latent inputs}}
    \label{fig:m52-sim-x-effects}
\end{figure}

\begin{figure}[!ht]
    \centering
    \includegraphics[width = \linewidth]{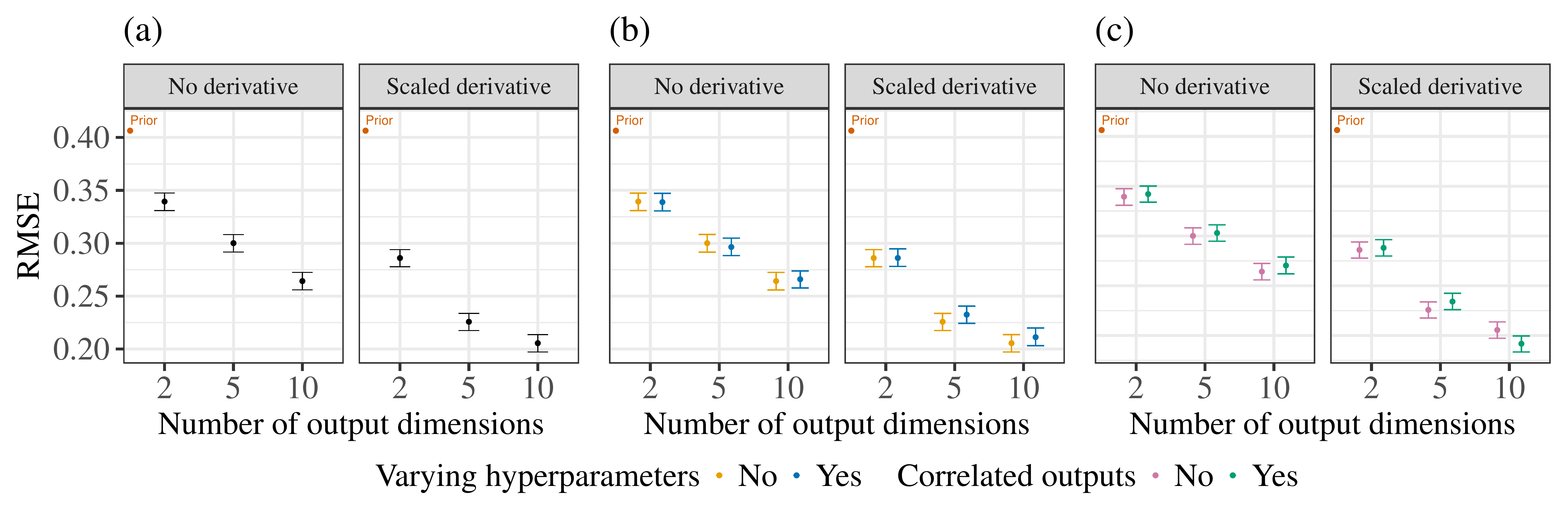}
    \caption{\textit{Periodic scenario: Main effects of including (a) scaled derivatives and interaction effects of assuming (b) varying hyperparameters and (c) correlated outputs on recovery of latent inputs}}
    \label{fig:per-sim-x-effects}
\end{figure}

Our findings for the latent $x$ are presented for different simulation scenarios in Figures \ref{fig:se-sim-x-effects}--\ref{fig:m52-sim-x-effects} for the simulated GP data scenarios and in Figure \ref{fig:per-sim-x-effects} for the periodic data scenario. We see how the inclusion of both derivative information and scaling modifications simultaneously results in an overall substantial decrease in mean RMSE in the simulated SE and periodic data scenarios (Figure \ref{fig:se-sim-x-effects}(a) and \ref{fig:per-sim-x-effects}(a)), thus indicating a better recovery of the true latent values as compared to models without derivative information. In case of the Matern 3/2 and 5/2 data scenarios, although not as substantial as the SE and periodic case, we see similar effects of adding scaled derivatives. This is due to the more challenging nature of the GPs with derivative Matern covariance functions as seen through their model convergence. Additionally, the evaluation measures for all the scenarios further decrease as we increase the number of output dimensions.

Overall, we see a reduction in RMSE of more than 50\% compared to the corresponding prior metrics, and a reduction of about 30\% compared to the models without derivative information in the SE and periodic data cases, thus clearly outlining the benefits of using DGP-LVMs. Conversely, when models include derivative information \textit{without} accounting for scale differences, the RMSEs are a lot higher, suggesting that the model performs adversely while estimating latent inputs (see Figures \ref{fig:se-sim-x-effects-rmse-full}--\ref{fig:per-sim-x-effects-rmse-full} in Appendix B). Curiously, the performance of such models is even worse than the models not including derivatives at all, sometimes close to (or even worse than) when just using the prior measurement model alone. Presumably, this is because hyperparameter estimates are strongly biased if forced to be the same for both regular outputs and their derivatives; at least when the ground truth assumes hyperparameters to be different by a factor of 3 (which is not unrealistic). As an implication, we then also obtain strongly biased latent input estimates, resulting in large RMSEs. This clearly highlights the importance of our derivative covariance function modifications. Without these modifications, using derivative information poses the risk of providing strongly misleading results.

With respect to the other varied components, modelling varying hyperparameters and correlated outputs may result in a slight increase in the RMSEs (see (b) and (c) of Figures \ref{fig:m32-sim-x-effects}--\ref{fig:per-sim-x-effects}), especially in higher output dimensions. We hypothesize that this is due to the significant increase in the number of estimated parameters, while the amount of data points remained constant in our simulations. Concretely, the number of parameters increase by the number of hyperparameters per dimension (i.e., $5$ in our case) times the number of output dimensions $D$, which is quite substantial already for $D=10$ output dimensions. 

We encounter a similar issue when modelling outputs as correlated since the increase in estimated model parameters are even more substantial. For output dimension $D$, we estimate $D(D-1)/2$ number of parameters just for between-dimension correlations. Such a significant increase in parameters becomes visible in the results especially for $D = 10$ for most of the simulated scenarios. 
\FloatBarrier
\subsubsection{Model evaluation: Hyperparameters}
\begin{figure}[!ht]
    \centering
    \includegraphics[width = \linewidth]{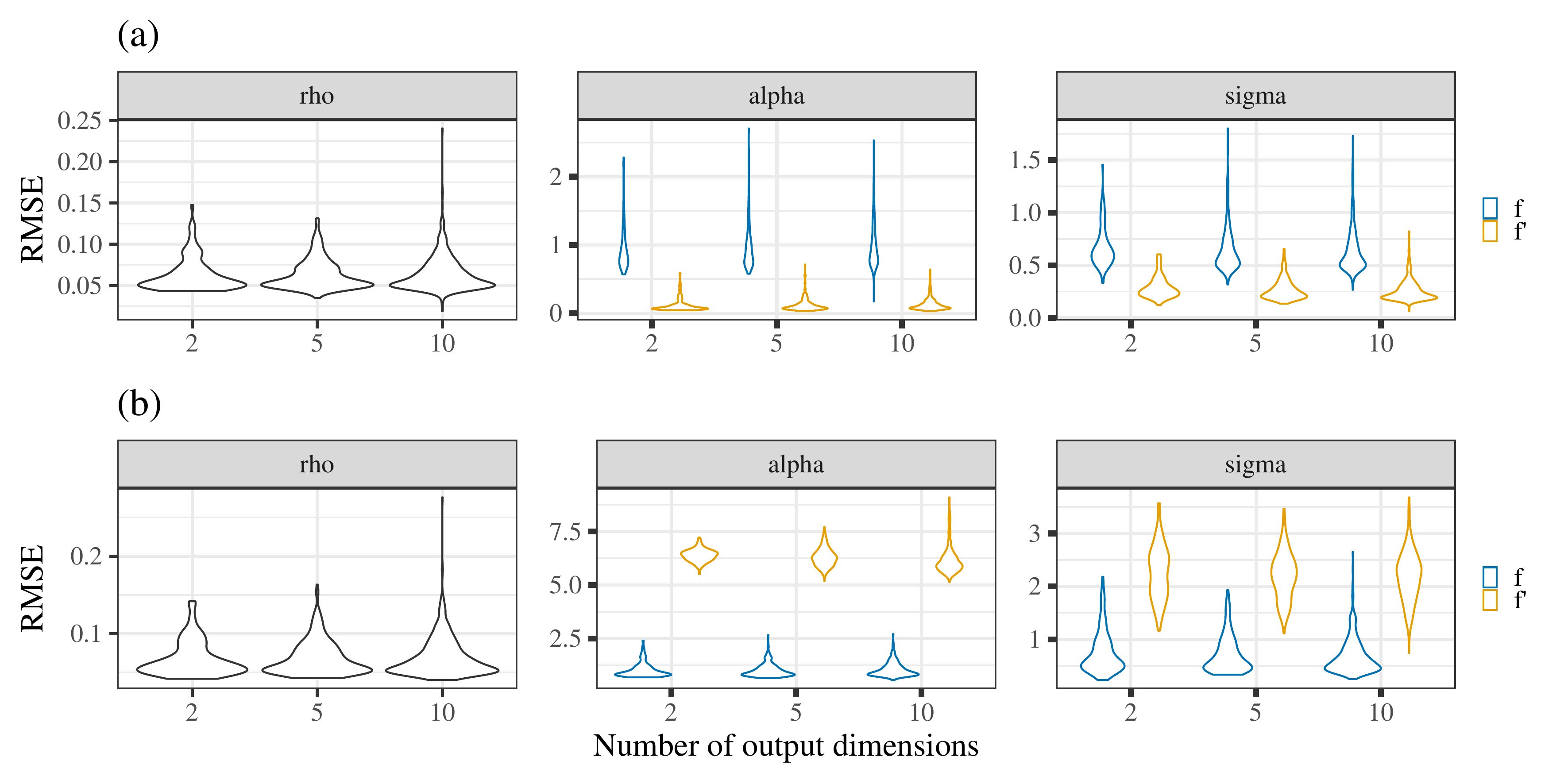}
    \caption{\textit{Squared exponential scenario: Hyperparameter RMSEs for (a) full DGP-LVM model and (b) models without scale assumption. The different colour denotes if the hyperparameters correspond to the original or the derivative part of the model.}}
    \label{fig:se-sim-hyperparams}
\end{figure}
\begin{figure}[!ht]
    \centering
    \includegraphics[width = \linewidth]{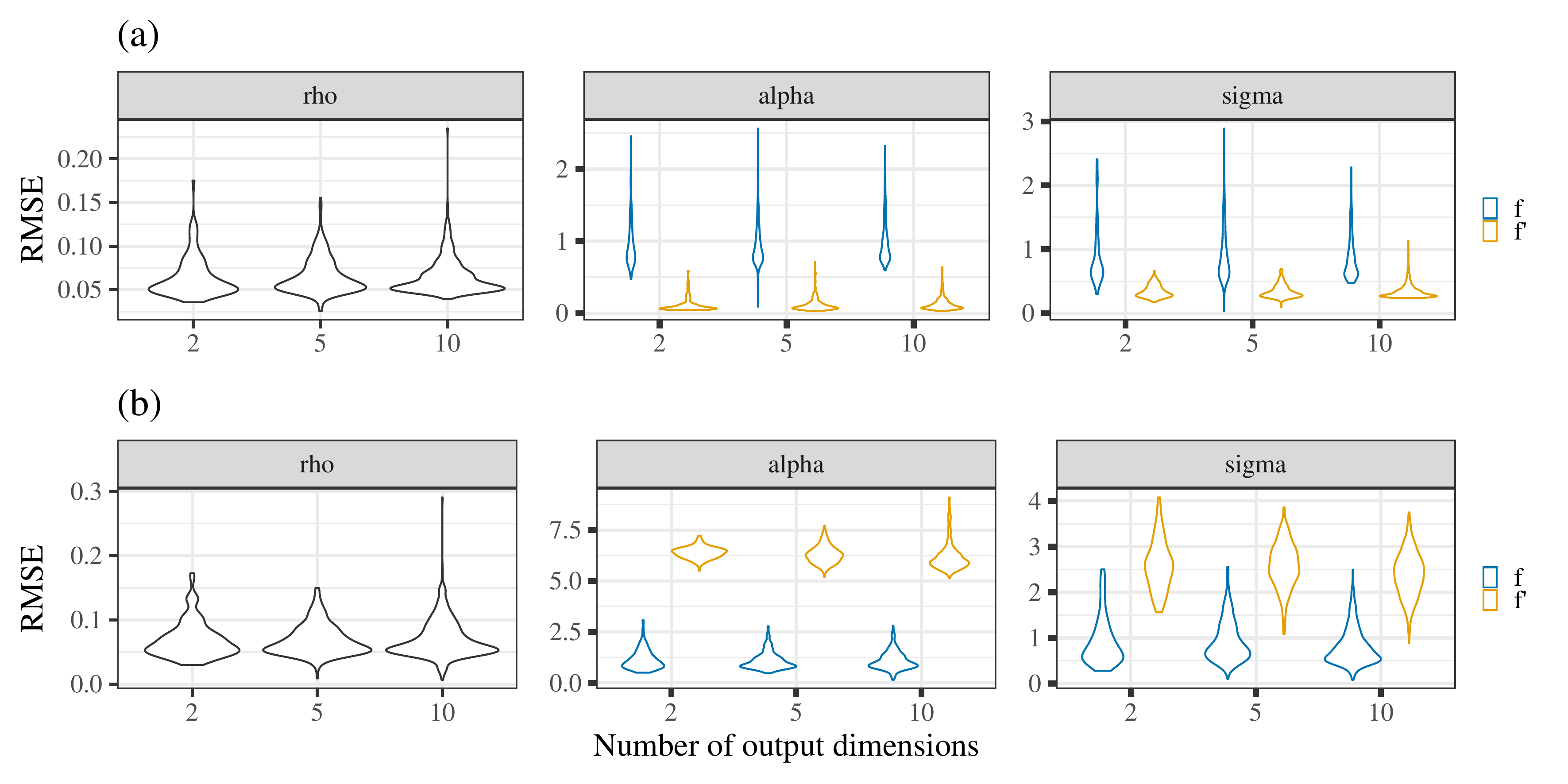}
    \caption{\textit{Matern 3/2 scenario: Hyperparameter RMSEs for (a) full DGP-LVM model and (b) models without scale assumption. The different colour denotes if the hyperparameters correspond to the original or the derivative part of the model.}}
    \label{fig:m32-sim-study-hyperparams}
\end{figure}
\begin{figure}[!ht]
    \centering
    \includegraphics[width = \linewidth]{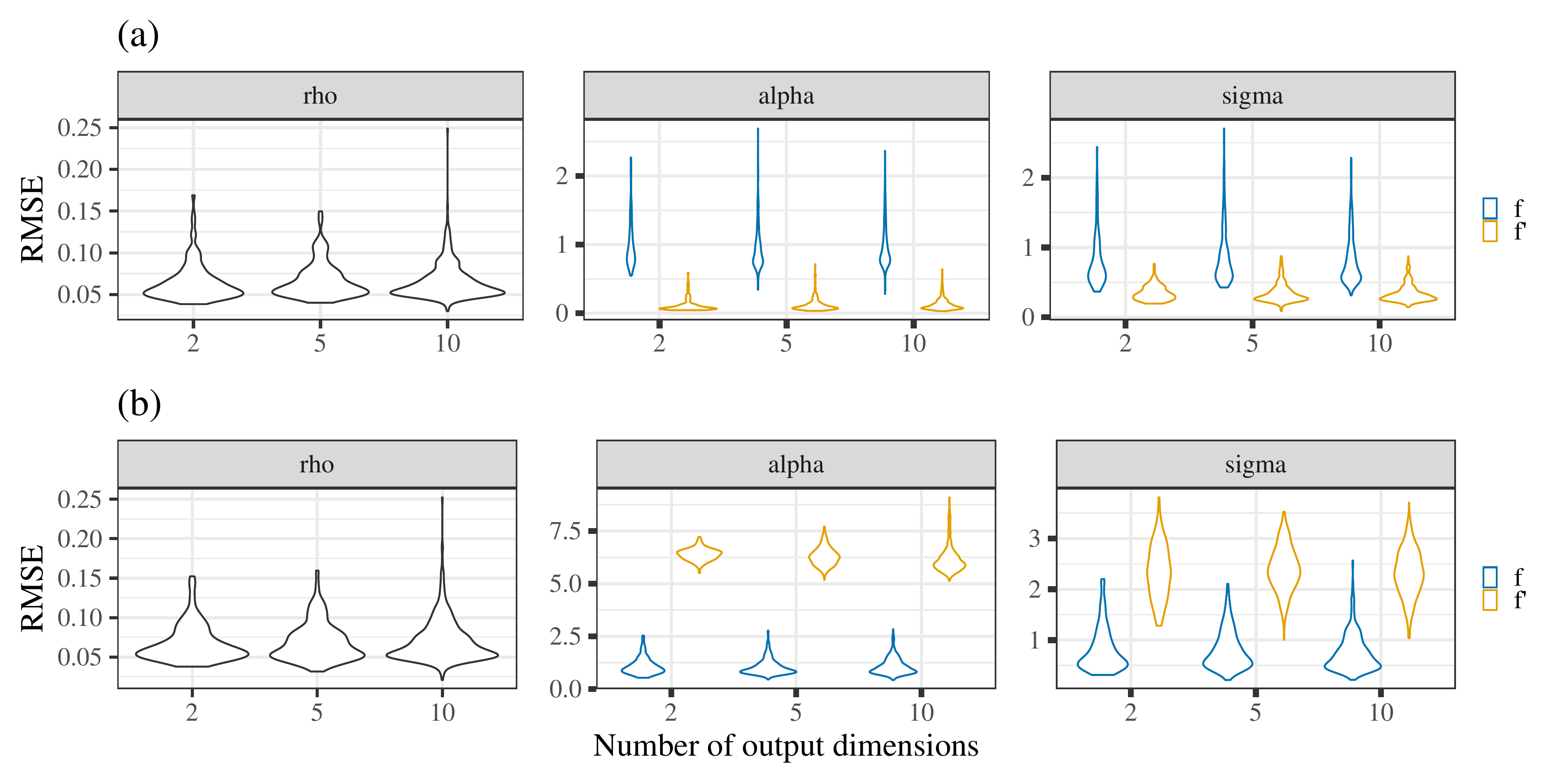}
    \caption{\textit{Matern 5/2 scenario: Hyperparameter RMSEs for (a) full DGP-LVM model and (b) models without scale assumption. The different colour denotes if the hyperparameters correspond to the original or the derivative part of the model.}}
    \label{fig:m52-sim-study-hyperparams}
\end{figure}
\begin{figure}[!ht]
    \centering
    \includegraphics[width = \linewidth]{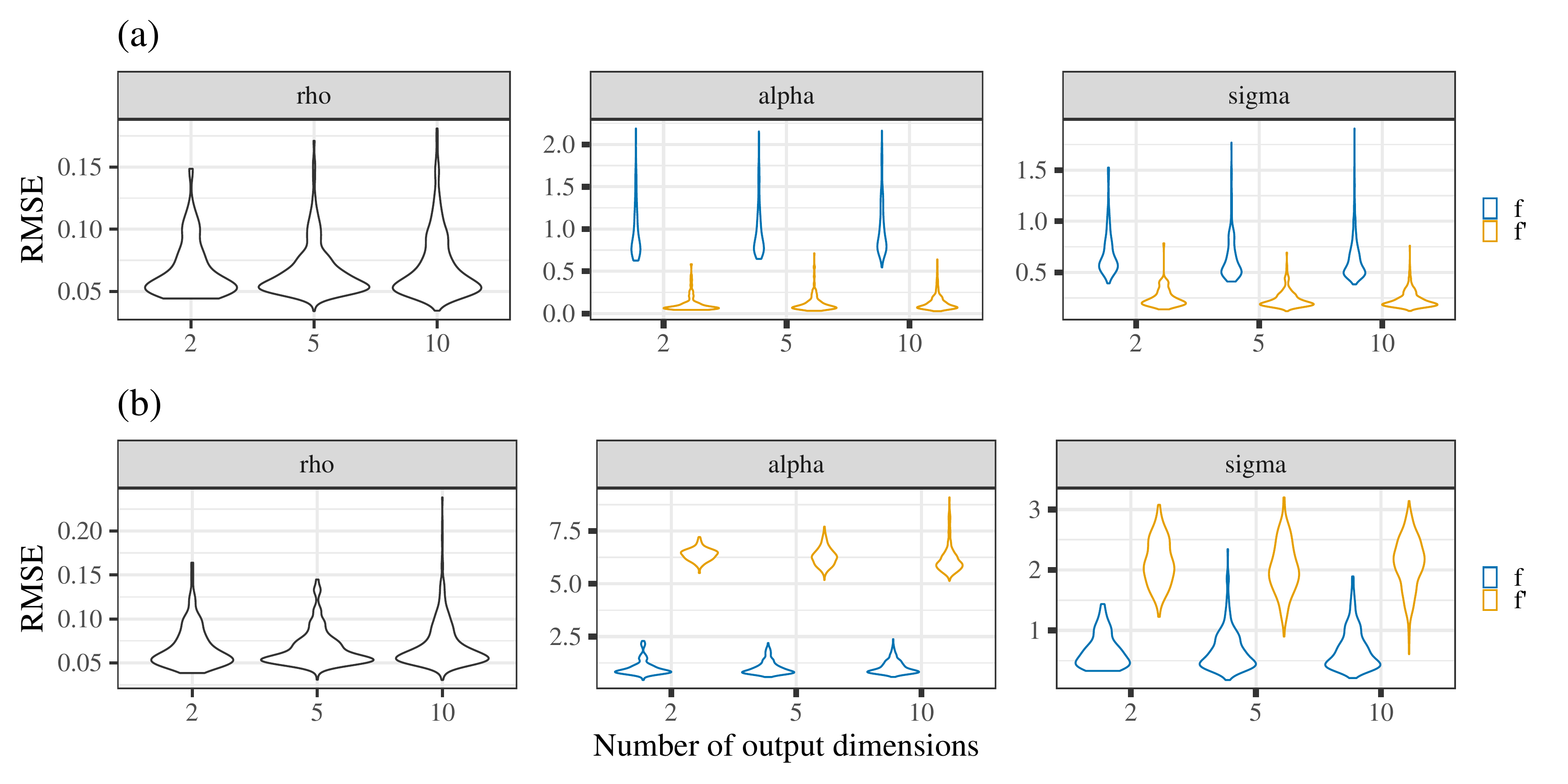}
    \caption{\textit{Periodic scenario: Hyperparameter RMSEs for (a) full DGP-LVM model and (b) models without scale assumption. The different colour denotes if the hyperparameters correspond to the original or the derivative part of the model.}}
    \label{fig:per-sim-study-hyperparams}
\end{figure}

In Figures \ref{fig:se-sim-hyperparams}--\ref{fig:per-sim-study-hyperparams}, we show hyperparameter recovery for the full DGP-LVM in the GP simulation scenarios and the periodic simulation scenario and compare the effects of accounting for scaling (each figure shows (a) full model with scaling and (b) without scaling). 

For most of the simulated datasets, the hyperparameters show good recovery as indicated by low RMSE. We do see some extreme RMSE values though, especially for GP length-scale $\rho$. These extreme cases are explained by the significant increase in the number of estimated parameters, when we consider both varying hyperparameters and correlated outputs without increasing the amount of data. 

Interestingly, we see how the scaling assumption helps identifying the GP marginal SD $\alpha'$ and error SD $\sigma'$ for the $f'$ and $y'$ respectively. Without the assumption, the model simply fails to recover the true SD hyperparameters for the derivative part of the data. Additionally, when disabling the assumptions of varying hyperparameters, correlated outputs separately as well as together (see (a), (b) and (c) of Figures \ref{fig:app-se-hyperparams-others}--\ref{fig:app-per-hyperparams-others} respectively in Appendix B), we see how recovery of hyperparameters, especially the GP marginal SDs struggle due to model misspecification. This demonstrates that each of the model innovations discussed in Section \ref{sec-methods} are important when the underlying data require such complexities.

\FloatBarrier
\subsubsection{Special case: Non-stationary data}

\begin{figure}[!ht]
    \centering
    \includegraphics[width = \linewidth]{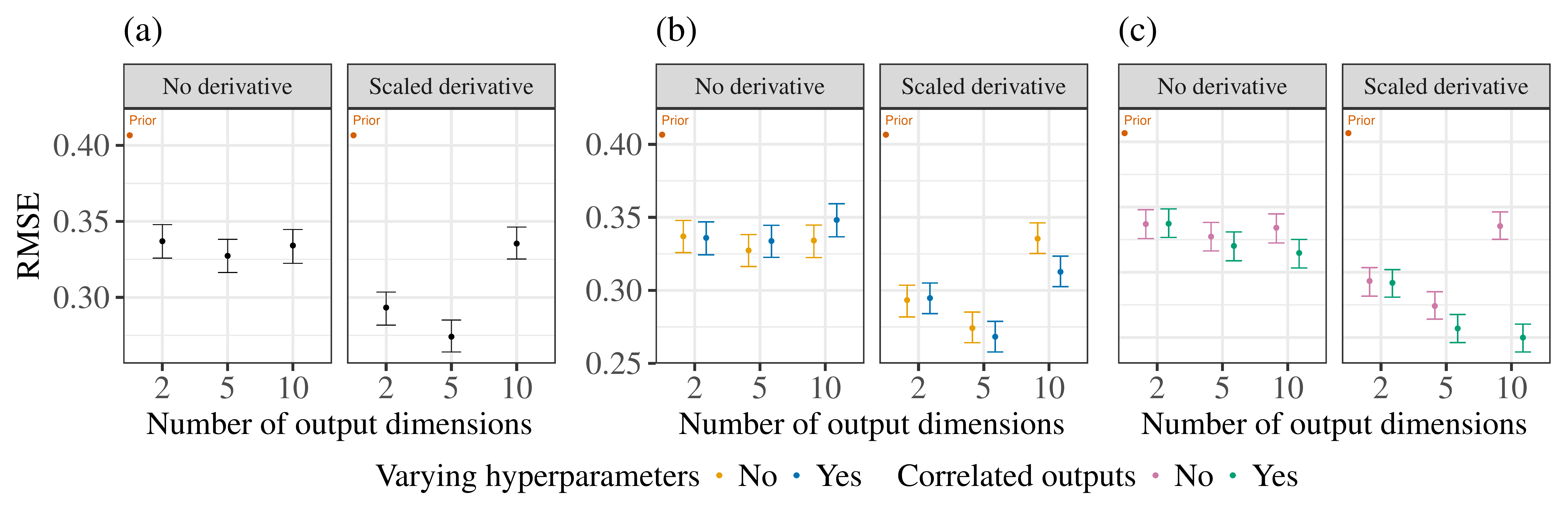}
    \caption{\textit{Periodic with trend scenario: Main effects of including (a) scaled derivatives and interaction effects of assuming (b) varying hyperparameters and (c) correlated outputs on recovery of latent inputs}}
    \label{fig:per-trend-sim-x-effects}
\end{figure}

In the additional case of periodic data with trend, Figure \ref{fig:per-trend-sim-x-effects} shows a sharp increase in RMSE values for higher dimensions (when $D = 10$).
We believe this to be a direct consequence of modelling non-stationary data with stationary GPs. With more number of non-stationary output dimensions, more of the stationary GPs fail to model the data appropriately, thus showing poor model performance in terms of recovering the ground truth of the latent $x$. This special simulation scenario highlights one of the limitations of our current framework, which we further discuss in Section \ref{sec-limit-future-res}. Interestingly, for the $D=10$, the problem vanishes when modelling correlated outputs. This is due to the fact that added trend behaviour is same across outputs owing to the shared inputs, thus being highly correlated to one another. Due to this, accounting for correlated outputs seem to improve the recovery of true latent inputs.

\section{Case study} \label{sec-case-study}

We showcase the application of DGP-LVM to real-world scRNA sequencing data by re-analysing cell-cycle data from \cite{mahdessian_spatiotemporal_2021}. This dataset comprises of single-cell RNA expression profiles along the cell cycle as well as corresponding RNA velocities as estimates of expression profile derivatives obtained as a pre-processing step of cytopath, a method for simulation based cell trajectory inference \citep{gupta_simulation-based_2022}. Briefly, we choose this dataset because it covers single-cell transcriptomic profiles of the cell cycle, i.e. a cyclic process going through four phases depicting substantial variation in gene expressions and velocities.

\begin{figure}[!ht]
    \centering
    \includegraphics[scale=0.4]{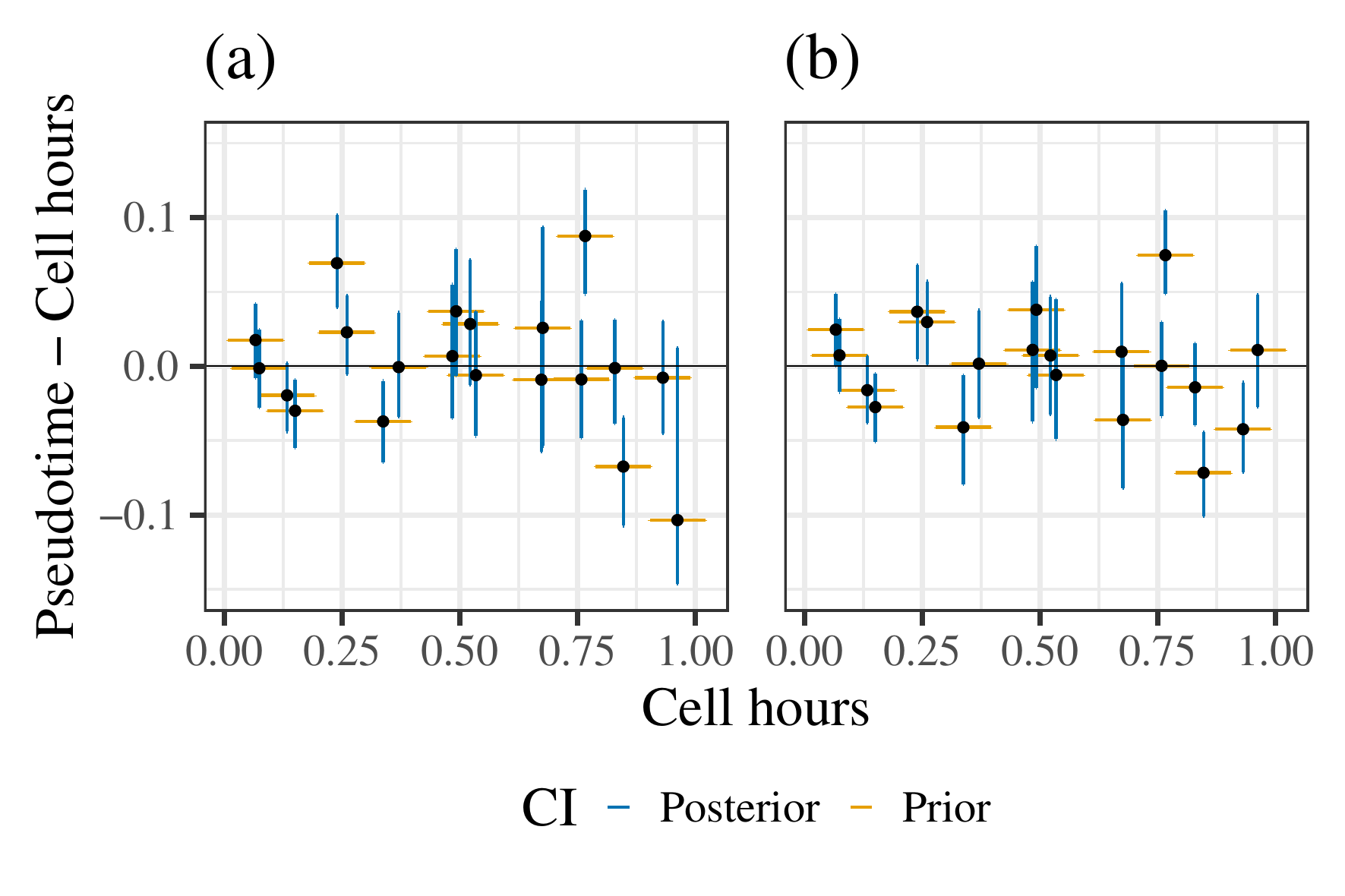}
    \caption{\textit{Difference of latent pseudotime estimates obtained via DGP-LVM with (a) SE and (b) Matern 5/2 covariance functions and prior cell hours. The point ranges horizontally show 95\% prior CIs and vertically show 95\% posterior CIs. Notice that the posterior CIs are actually much smaller than the prior CIs since y-axis scale is significantly smaller than x-axis.}}
    \label{fig:case-study-output}
\end{figure}

For the purpose of this case study, we use a reduced data set of spliced RNA gene expression data and its corresponding RNA velocity of 20 cells and 12 genes. In other words, each sample point corresponds to a single cell and each output dimension corresponds to a single gene, with the value being the gene expressions per cell. Thus, for this case study, we have the sample points $N = 20$ for $y$ and correspondingly $y'$ each with output dimensions $D = 12$. We sub-sampled the dataset in a stratified fashion so that cells from all four phases are included. We use the experimental time known as "cell hours" in the context of this specific data as the prior $\tilde{x}$ for our latent pseudotime (input) $x$. Both $\tilde{x}$ and $x$ are real numbers with values ranging between 0 and 1. For our prior measurement SD, we choose $s = 0.03$, so that it is proportional to our choices in simulation studies in Section \ref{sec-sim-data}.

We fit DGP-LVMs with derivative SE and Matern 5/2 covariance functions. For this case study, We specify the priors for GP marginal SDs $\alpha \sim \text{Normal}^+(13.84, 3.46^2)$ and $\alpha' \sim \text{Normal}^+(1, 0.25^2)$. In case of error SDs we specify $\sigma \sim \text{Normal}^+(6.92, 3.46^2)$ and $\sigma' \sim \text{Normal}^+(0.5, 0.25^2)$. These choices were influenced by the large scaling factor $\lambda$ informed by the data, as the mean and standard deviations of $y$ are on average 13.84 times larger than those of $y'$ across the output dimensions. The prior for $\rho$ was specified as $\text{Normal}^+(0.4, 0.1^2)$ for the SE and $\text{Normal}^+(0.6, 0.1^2)$ for the Matern 5/2 as our GP length scale priors loosely based on the scale of the latent input $x$ (as suggested by the prior $\tilde{x}$). These priors were chosen to account for the varying functional smoothness induced by the choice of covariance function. The DGP-LVM with Matern 3/2 being the least functionally smooth choice of covariance function from the Matern family didn't converge reasonably with a sensible choice of $\rho$ prior for this specific data and is therefore not presented.

\begin{figure}[!ht]
    \centering
    \includegraphics[width = \linewidth]{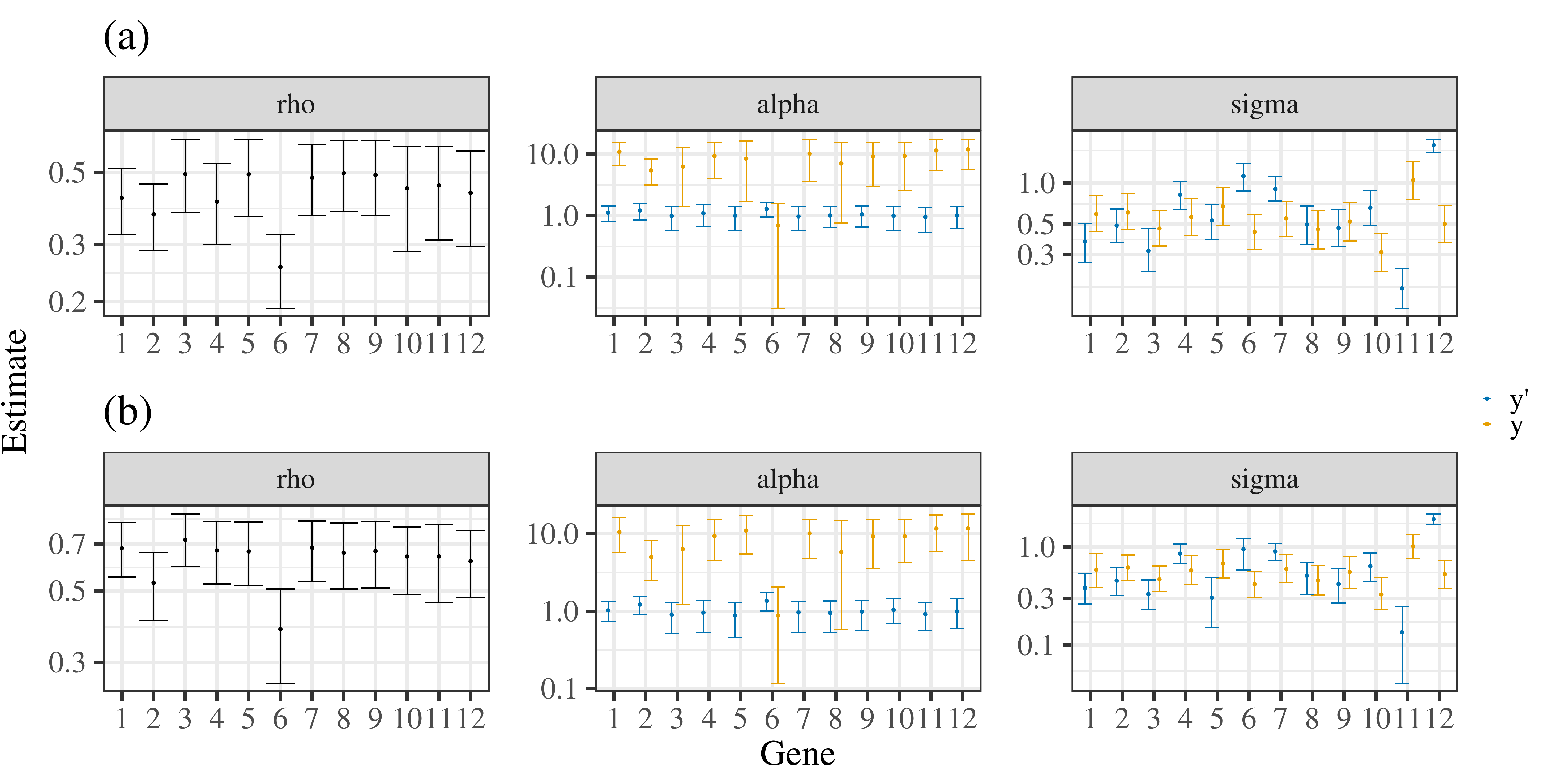}
    \caption{\textit{Hyperparameters for DGP-LVM with (a) SE and (b) Matern 5/2 covariance functions. The points indicate posterior mean and the point ranges indicate 95\% CIs for each hyperparameter per output dimension. The different colours denote correspondence to the output $y$ or its derivative $y'$.}}
    \label{fig:case-study-hyperparams}
\end{figure}

As in any real-world latent variable estimation problem, we lack the ground truth to compare the estimated latent values against. Therefore, we study the deviation of the posterior estimates of pseudotime from the cell hours (our prior) by considering the difference or shift in values of the estimated pseudotime from the observed cell hours. The results are shown in Figure \ref{fig:case-study-output} with cell hours (prior) on the x-axis and shift (difference of pseudotime and cell hours) on the y-axis. Deviations from the $y=0$ line indicate that latent pseudotimes are different from their cell hours (prior) as a result of learning from gene expression data and velocities. For some cells, prior-posterior differences are up to 5\% of the total time scale. Further, we see that the posterior uncertainties (error bars in y-direction) are substantially smaller (considering the scale of the y-axis compared to x-axis) than the corresponding prior uncertainties (error bars in x-direction), which also indicates that model learning has taken place. Combined with our findings from the simulation study as strong evidence that DGP-LVM is able to learn and recover the true posterior estimates of the latent pseudotimes, these deviations from the prior are interpreted as model learning in the correct direction closer to the true latent ordering of the cells.

In Figure \ref{fig:case-study-hyperparams}, we show the posterior mean along with SD estimates of GP hyperparameters. We see strongly varying length-scales $\rho$, marginal SDs $\alpha$ and error SDs $\sigma$ across different genes (output dimensions) for both SE and Matern 5/2 models. This clearly points to the necessity of modelling hyperparameters as varying across genes. We also see substantial scale differences between the GP marginal SDs $\alpha$ and $\alpha'$ corresponding to $f$ and $f'$, and consequently gene expression $y$ and velocity $y'$ outputs, respectively. Similar, but not as drastic results are seen for the error SDs $\sigma$ and $\sigma'$. These results indicate significant scale differences between output RNA gene expression and its derivative RNA velocity. Interestingly, the scale differences go in both directions, such that for some genes $\alpha$ and $\sigma$ are higher than $\alpha'$ and $\sigma'$. For others, the direction is opposite. While the direction is not important for the DGP-LVM models, it may be highly relevant for understanding the biological processes in which the specific genes are involved. The difference in posterior estimates of the hyperparameters in the different models are heavily influenced by the natural varying functional smoothness corresponding to the choice of different covariance functions.

That said, this case study is meant only as a simple example for demonstrating the application of DGP-LVMs on real-world data. We would like to caution against any specific biological interpretation of the results at this point. The case study was conducted using Apple M1 chip (2 cores in parallel) with 16 GB memory allowances. The runtime (in minutes) for DGP-LVM were approximately 25.3 for the SE model and 73.11 for the Matern 5/2 model.

\section{Discussion} \label{sec-discuss}

Motivated by a real-world problem in the area of single-cell biology, we developed a class of derivative Gaussian process latent variable models, DGP-LVMs. In the real-world case, we aim at estimating the latent ordering of cells from RNA gene expression levels and its corresponding time derivative RNA velocity. For this purpose, DGP-LVMs not only account for scale differences between the outputs and their derivatives, but also learn from multiple, potentially correlated outputs. The latter is highly important for scRNA sequencing data where the latent cell ordering is informed by many genes, each forming their own output coupled with derivative information. 

In our simulation studies, we extensively validate DGP-LVMs demonstrating strong improvements in estimation accuracy of the latent variables by including derivative information. Our results also clearly show the importance of our proposed covariance function modifications. While we specifically focused on modifying the SE and Matern class of covariance functions, our framework is generally applicable for any choice of covariance function that is twice differentiable. 

\subsection{Limitations and Future Research}\label{sec-limit-future-res}

This paper is only the first step towards tackling latent variable (input) estimation with derivative Gaussian processes. The current main limitation of DGP-LVMs is their data-scalability as they cannot be easily applied to large amounts of data, such as full sized scRNA sequencing datasets, yet. For a dataset of size $N$, exact GPs have a complexity of $O(N^3)$ in operations and $O(N^2)$ in memory. In case of multi-output GPs with correlated outputs and varying hyperparameters, the complexities increase to $O(N^3 D^3)$ and $O(N^2 D^2)$ respectively, where $D$ is the number of output dimensions \citep{HensmanFL13}. Additionally, when performing Bayesian inference via HMC involving a total of $T$ unnormalized log posterior evaluations, the number of operations increases to even $O(N^3 D^3 \, T)$. Together, this limits inference for exact GPs on data with large $N$ or $D$. From a real-world data point of view, scRNA sequencing data frequently has a few thousand cells (sample size $N$) with the number of genes (output dimensions $D$) being in the high hundreds after standard pre-processing steps. In case of DGP-LVM, this issue is even more severe due to adding derivative information, effectively doubling the sample size $N$. To address the computational limitations of DGP-LVM in terms of data-scalability, future research should consider extending approximate GP approaches \citep[e.g.,][]{riutort-mayol_approx_gps_2022} to our DGP-LVM framework. 

Another limitation of our current DGP-LVMs is their stationary assumption based on the choice of the covariance functions we discuss here. This limits their applicability to non-stationary data as evidenced by our simulation study of periodic data with an added non-linear trend. While this is a general limitation of stationary GPs, the limitation currently lies in not having a derivative version of non-stationary covariance functions. An interesting future research would be to develop DGP-LVMs for non-stationary data where the primary focus would be on obtaining derivative versions of non-stationary covariance functions and verifying their performance for latent variable estimation.

Another aspect for future research is the choice of prior distributions. Here, we focused on informative priors for the GP hyperparameters in both our simulation studies and the real-world case study although they are difficult to come by organically. DGP-LVMs will likely benefit from using stronger priors informed by the application-specific subject matter knowledge, specifically in data-sparse scenarios. This not only applies to priors for the GP hyperparameters, but also to the priors of the latent input variables. Moreover, a joint prior on the covariance function hyperparameters along with latent inputs will likely further improve model convergence. The combination of scalable approximations, improved prior specifications, and additional derivative covariance functions would foster the general applicability of DGP-LVMs, thereby further increasing their ability to accurately estimate latent variables. 

\section*{Code availability}
The code for the model development, simulation studies as well as the results can be found here: \url{https://github.com/Soham6298/DGP-LVM}.

\section*{Acknowledgments}
We acknowledge the Cluster of Excellence iFIT (EXC 2180) "Image-Guided and Functionally Instructed Tumor Therapies" for supporting Soham Mukherjee. We acknowledge the valuable insights provided by members of ClaassenLab and BürknerLab, Revant Gupta and Marcello Zago. We thank Debapratim Sil and Jayati Chatterjee for their feedback on the manuscript. The authors thank the International Max Planck Research School for Intelligent Systems (IMPRS-IS) for supporting Soham Mukherjee.

\bibliographystyle{apalike}
\bibliography{refs} 

\section*{Appendix} \label{App}

\subsection*{Appendix A: Derivative covariance functions} \label{app-math}

Here we show the mathematical details of the general covariance function structure as well as specific derivative forms of SE, Matern 3/2 and Matern 5/2 covariance functions.

\subsubsection*{Proof of the derivative covariance function structure} \label{App-deriv-covfn-proof}

\textbf{Lemma}: Let $X \in \mathcal{X}$ be a random variable and $g:\mathbb{R} \times \mathcal{X} \rightarrow \mathbb{R}$ is a function $\ni g(t,X)$ is integrable $\forall t$ and $g$ is continuously differentiable w.r.t $t$. Assume a random variable $Z \ni \left|\frac{\delta}{\delta t}g(t,X)\right| \le Z$ almost surely $\forall t$ and $E(Z)<\infty$, then $\frac{\delta}{\delta t} E(g(t,X)) = E\left(\frac{\delta}{\delta t} g(t,X) \right)$. 

Let us consider $y_i$ and $v_j$ such that $v_j = \frac{\delta y_j}{\delta x_j}$ where we consider a Gaussian Process model with $y = f(x) + \epsilon$. For a GP model $\text{Cov}(y_i,y_j)$ is completely defined using corresponding inputs $x_i$ and $x_j$ by a covariance function.
We see that
\begin{flalign*}
    \text{Cov}(y_i, v_j) &= \mathbb{E}\left(\left(y_i - \mathbb{E}\left(y_i\right)\right)\left(v_j - \mathbb{E}(v_j)\right) \right) &\\
    &= \mathbb{E}\left(\left(y_i - \mathbb{E}\left(y_i\right)\right)\left(\frac{\delta}{\delta x_j} y_j - \mathbb{E}\left(\frac{\delta}{\delta x_j} y_j \right) \right) \right) &\\
    &= \mathbb{E}\left(\left(y_i - \mathbb{E}\left(y_i\right)\right)\left(\frac{\delta}{\delta x_j} y_j - \frac{\delta}{\delta x_j}\left(\mathbb{E}\left( y_j \right) \right)\right)\right) && \text{(By DCT)} \\
    &= \mathbb{E}\left(\left(y_i - \mathbb{E}\left(y_i\right)\right)\frac{\delta}{\delta x_j} \left(y_j - \left(\mathbb{E}\left( y_j \right) \right)\right)\right) && \text {(derivative over subtraction)}\\
    &= \mathbb{E}\left(\frac{\delta}{\delta x_j}\left(y_i - \mathbb{E}\left(y_i\right)\right) \left(y_j - \left(\mathbb{E}\left( y_j \right) \right)\right)\right) && \text {($y_i$'s are constant w.r.t $x_j$)}\\
    &= \frac{\delta}{\delta x_j}\left(\mathbb{E}\left(\left(y_i - \mathbb{E}\left(y_i\right)\right) \left(y_j - \left(\mathbb{E}\left( y_j \right) \right)\right) \right) \right)&& \text {(By DCT)}\\
    &= \frac{\delta}{\delta x_j} \text{Cov}(y_i, y_j).
\end{flalign*} 

Using similar reasoning and with $v_i = \frac{\delta y_i}{\delta x_j}$ and $v_j = \frac{\delta y_j}{\delta x_j}$, we find
\begin{flalign*}
   \text{Cov}(v_i,v_j) &= \mathbb{E}\left(\left(v_i - \mathbb{E}\left(v_i\right)\right)\left(v_j - \mathbb{E}\left(v_j\right)\right) \right) &\\
    &= \mathbb{E}\left(\left(\frac{\delta}{\delta x_i}y_i - \mathbb{E}\left(\frac{\delta}{\delta x_i}y_i\right)\right)\left(\frac{\delta}{\delta x_j} y_j - \mathbb{E}\left(\frac{\delta}{\delta x_j} y_j \right) \right) \right) &\\
    &= \mathbb{E}\left(\left(\frac{\delta}{\delta x_i}y_i - \frac{\delta}{\delta x_i} \mathbb{E}\left(y_i\right)\right)\left(\frac{\delta}{\delta x_j} y_j - \frac{\delta}{\delta x_j} \mathbb{E}\left(y_j \right) \right) \right) && \text{(By DCT)}\\
    &= \mathbb{E}\left(\frac{\delta}{\delta x_i}\left(\frac{\delta}{\delta x_j}\left(y_i - \mathbb{E}\left(y_i\right)\right)\left(y_j - \mathbb{E}\left(y_j\right)\right) \right) \right) && \text{(derivative over constants)}\\
    &= \frac{\delta^2}{\delta x_i \delta x_j}\text{Cov}(y_i, y_j) && \text{(By DCT)}
\end{flalign*}

\subsubsection*{Derivative covariance functions} \label{App-cov-fns}
In the following covariance functions, $\alpha$ is GP marginal SD corresponding to output $y$; $\alpha'$ is the GP marginal SD corresponding to derivative output $y'$; $\rho$ is the GP length scale parameter.

\begin{enumerate}
    \item \textbf{Squared Exponential}
        \begin{flalign*}
        K &= \alpha^2 \exp \left(-\frac{(x_i - x_j)^2}{2\rho^2}\right) &\\ \\
        K_{01} &= \frac{\delta K}{\delta x_j}= \alpha\alpha' \frac{(x_i - x_j)}{\rho^2} \exp \left(-\frac{(x_i - x_j)^2}{2\rho^2}\right) &\\ \\
        K_{10} &= \frac{\delta K}{\delta x_i} = \alpha\alpha' \frac{(x_j - x_i)}{\rho^2} \exp \left(-\frac{(x_i - x_j)^2}{2\rho^2}\right) &\\ \\
        K_{11} &= \frac{\delta^2 K}{\delta x_i \delta x_j} = \frac{\alpha'^2}{\rho^4}(\rho^2 - (x_i - x_j)^2) \exp\left(-\frac{(x_i - x_j)^2}{2\rho^2}\right) &
        \end{flalign*}
    \item \textbf{Matern 3/2}
        \begin{flalign*}
        K &= \alpha^2 \left(1 + \frac{\sqrt{3(x_i - x_j)^2}}{\rho} \right) \exp\left(-\frac{\sqrt{3(x_i - x_j)^2}}{\rho}\right) &\\ \\
        K_{01} &= \frac{\delta K}{\delta x_j} = \alpha\alpha' \left( \frac{3(x_i-x_j)}{\rho^2} \right)\exp\left(-\frac{\sqrt{3(x_i - x_j)^2}}{\rho}\right) &\\ \\
        K_{10} &= \frac{\delta K}{\delta x_i} = \alpha\alpha' \left( \frac{3(x_j-x_i)}{\rho^2} \right)\exp\left(-\frac{\sqrt{3(x_i - x_j)^2}}{\rho}\right) &\\ \\
        K_{11} &= \frac{\delta^2 K}{\delta x_i \delta x_j} = \alpha'^2\left(\frac{3}{\rho^2}\right) \left(1 - \frac{\sqrt{3(x_i-x_j)^2}}{\rho} \right) \exp\left(-\frac{\sqrt{3(x_i - x_j)^2}}{\rho}\right) &
        \end{flalign*}
    \item \textbf{Matern 5/2}
        \begin{flalign*}
        K &= \alpha^2 \left(1 + \frac{\sqrt{5(x_i - x_j)^2}}{\rho} + \frac{5(x_i-x_j)^2}{3\rho^2} \right) \exp\left(-\frac{\sqrt{5(x_i - x_j)^2}}{\rho}\right) &\\ \\
        K_{01} &= \frac{\delta K}{\delta x_j} = \alpha\alpha' \left( \frac{5(x_i-x_j)}{3\rho^2} \right) \left(1 + \frac{\sqrt{5(x_i-x_j)^2}}{\rho} \right) \exp\left(-\frac{\sqrt{5(x_i - x_j)^2}}{\rho}\right) &\\ \\
        K_{10} &= \frac{\delta K}{\delta x_i} = \alpha\alpha' \left( \frac{5(x_j-x_i)}{3\rho^2} \right) \left(1 + \frac{\sqrt{5(x_i-x_j)^2}}{\rho} \right) \exp\left(-\frac{\sqrt{5(x_i - x_j)^2}}{\rho}\right) &\\ \\
        K_{11} &= \frac{\delta^2 K}{\delta x_i \delta x_j} = \alpha'^2\left(\frac{5}{3\rho^2}\right) \left(1 + \frac{\sqrt{5(x_i-x_j)^2}}{\rho} - \frac{5(x_i-x_j)^2}{\rho^2} \right) \exp\left(-\frac{\sqrt{5(x_i - x_j)^2}}{\rho}\right) 
        &
        \end{flalign*}
\end{enumerate}

\newpage

\subsection*{Appendix B: Additional simulation results} \label{app-plots}
Here we show the additional plots from our simulation studies. Specifically, we provide the MCMC convergence diagnostics for Matern 3/2, 5/2 as well as the periodic simulation scenarios. We then present the full versions of the model evaluation plots for recovery of true latent inputs $x$ using RMSE and MAE as evaluation metrics for all the simulation scenarios. Further we show additional hyperparameter recovery plots based on enabling/disabling different model assumptions. Finally, we present the case where we use four MCMC chains for a reduced SE data simulation scenario that shows minimal or no effect on model evaluation metrics irrespective of number of MCMC chains.
\FloatBarrier
\subsubsection*{Additional MCMC diagnostic plots} \label{app-mcmc-diag}

\begin{figure}[!htb]
    \centering
    \includegraphics[width = \linewidth]{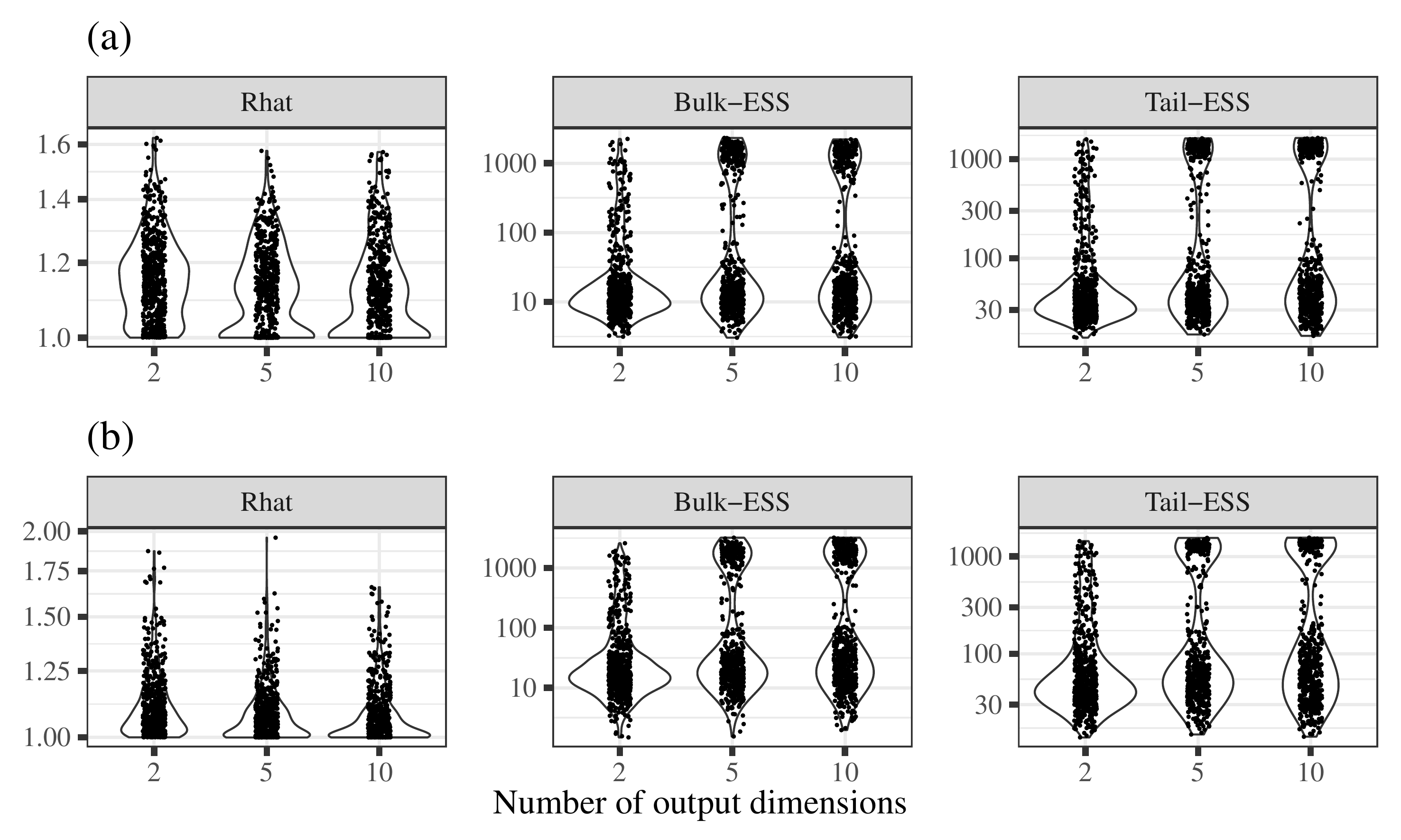}
    \caption{\textit{Matern 3/2 scenario: Convergence measures for (a) latent inputs and (b) GP hyperparameters. The individual points correspond to each fitted models per simulated data. The y-axis for Bulk and Tail ESS plots are log10 transformed.}}
    \label{fig:m32-model-validation-gp}
\end{figure}
\begin{figure}[!htb]
    \centering
    \includegraphics[width = \linewidth]{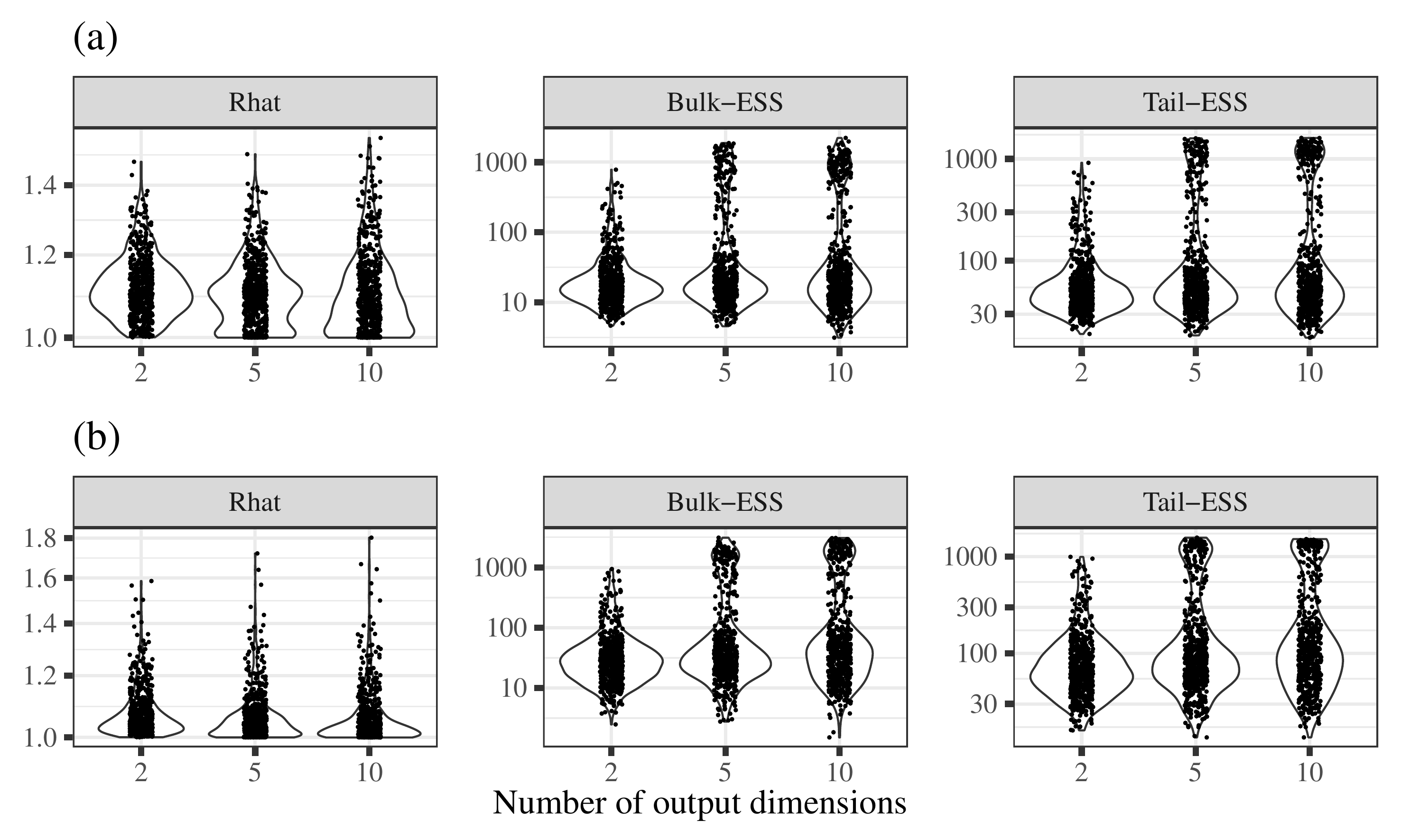}
    \caption{\textit{Matern 5/2 scenario: Convergence measures for (a) latent inputs and (b) GP hyperparameters. The individual points correspond to each fitted models per simulated data. The y-axis for Bulk and Tail ESS plots are log10 transformed.}}
    \label{fig:m52-model-validation-gp}
\end{figure}
\begin{figure}[!htb]
    \centering
    \includegraphics[width = \linewidth]{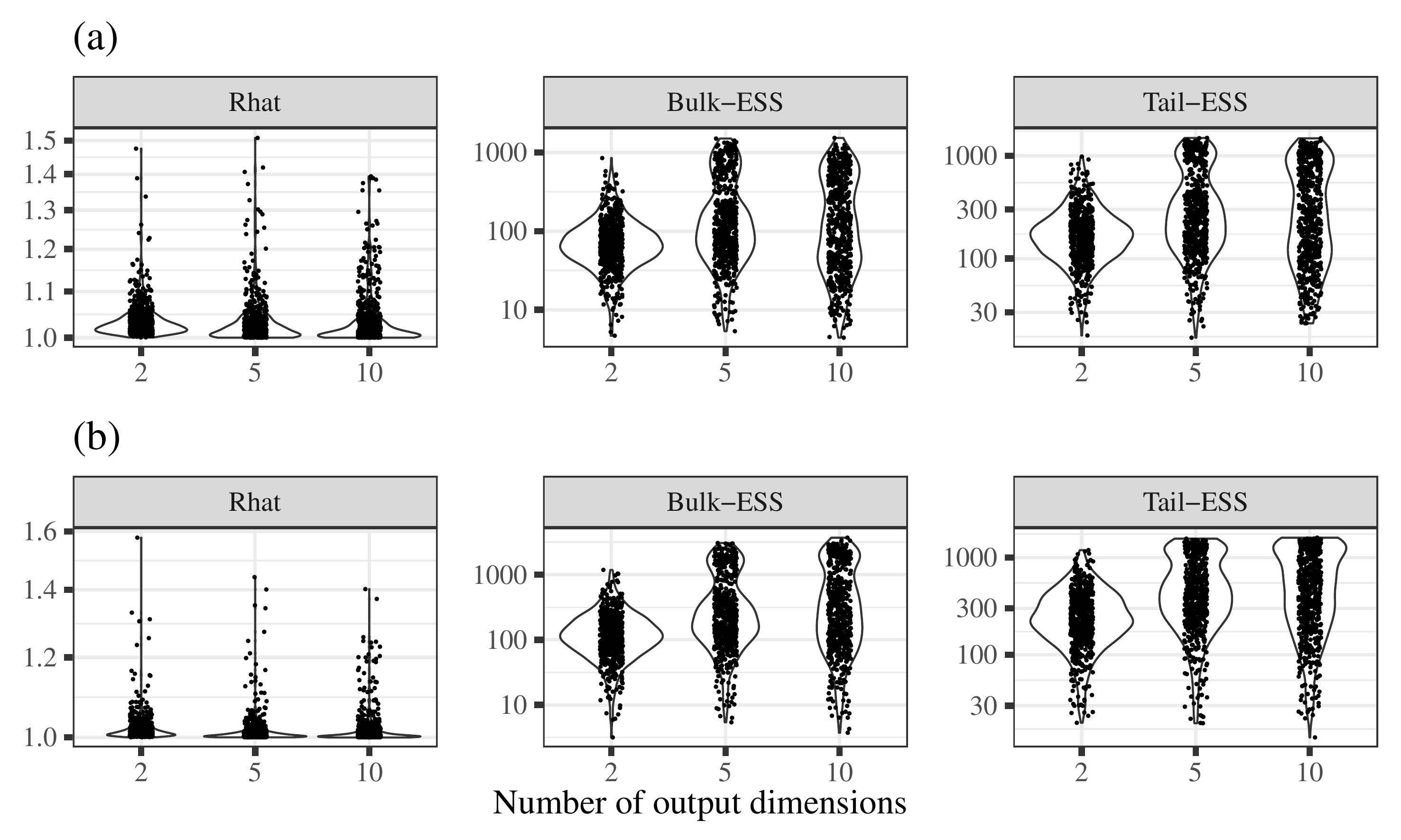}
    \caption{\textit{Periodic scenario: Convergence measures for (a) latent inputs and (b) GP hyperparameters. The individual points correspond to each fitted models per simulated data. The y-axis for Bulk and Tail ESS plots are log10 transformed.}}
    \label{fig:per-model-validation-gp}
\end{figure}
\begin{figure}[!htb]
    \centering
    \includegraphics[width = \linewidth]{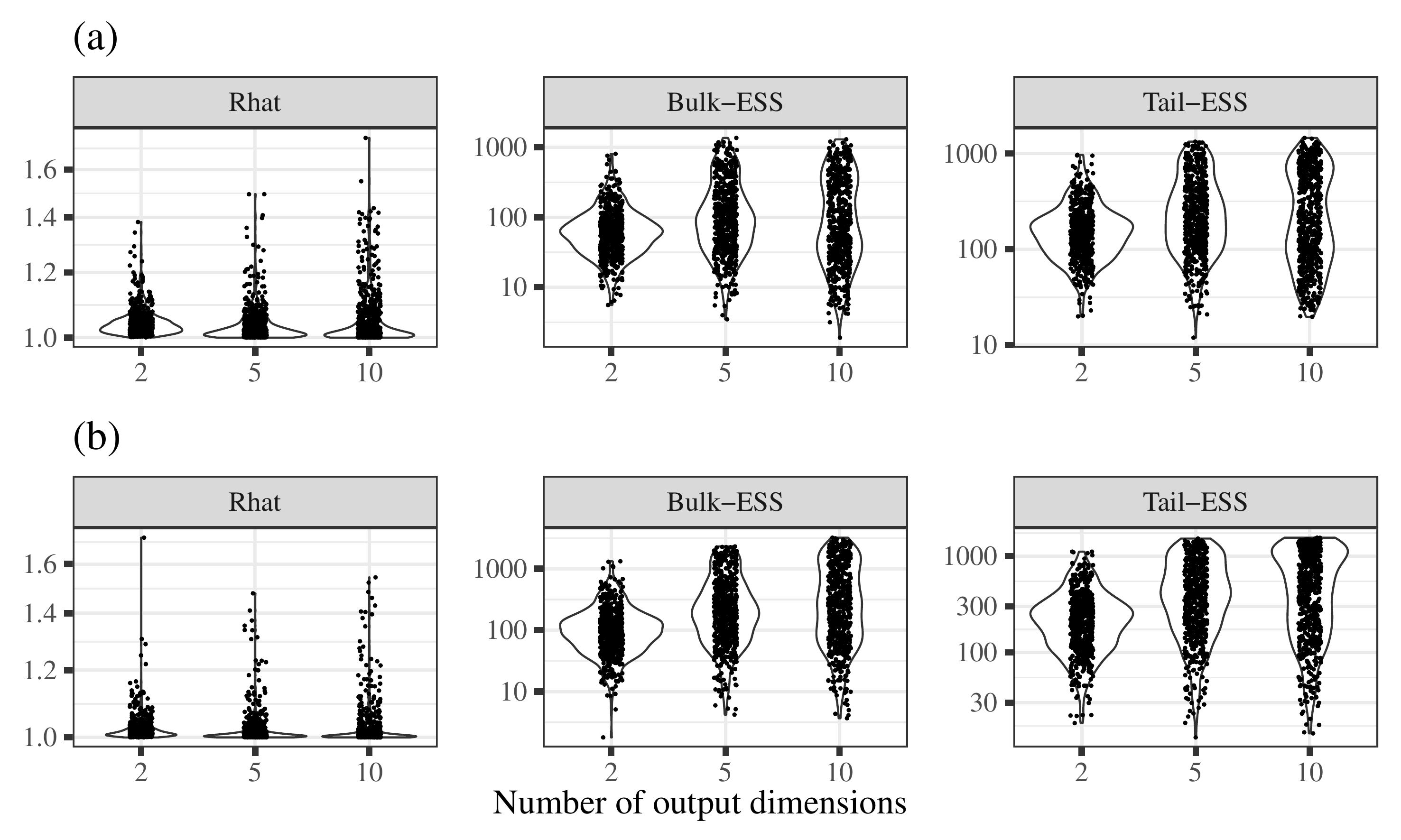}
    \caption{\textit{Periodic with trend scenario: Convergence measures for (a) latent inputs and (b) GP hyperparameters. The individual points correspond to each fitted models per simulated data. The y-axis for Bulk and Tail ESS plots are log10 transformed.}}
    \label{fig:per-trend-model-validation-gp}
\end{figure}
\newpage
\FloatBarrier
\subsubsection*{Full versions of model evaluation plots using RMSE: Latent inputs} \label{app-latent-eval-rmse}

\begin{figure}[!htb]
    \centering
    \includegraphics[width = \linewidth]{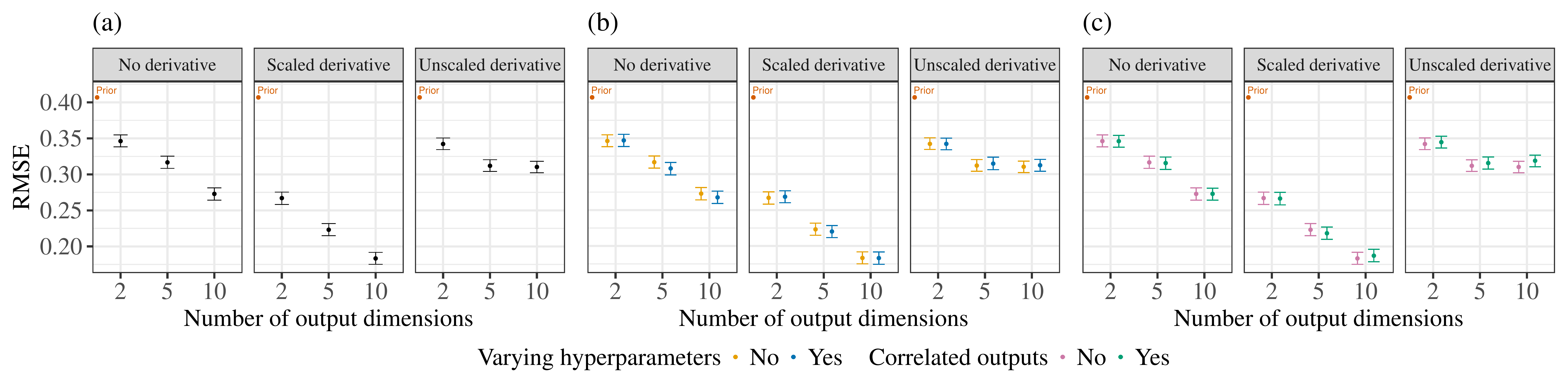}
    \caption{\textit{Squared exponential scenario: main effects of including (a) scaled derivatives and interaction effects of assuming (b) varying hyperparameters and (c) correlated outputs on recovery of latent inputs (full version) using RMSE.}}
    \label{fig:se-sim-x-effects-rmse-full}
\end{figure}

\begin{figure}[!htb]
    \centering
    \includegraphics[width = \linewidth]{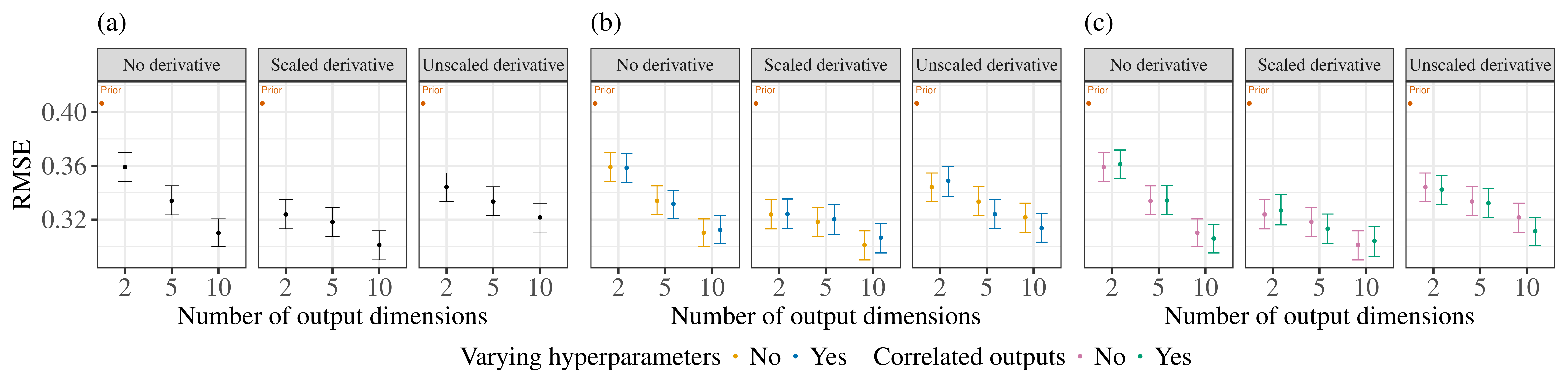}
    \caption{\textit{Matern 3/2 scenario: main effects of including (a) scaled derivatives and interaction effects of assuming (b) varying hyperparameters and (c) correlated outputs on recovery of latent inputs (full version) using RMSE.}}
    \label{fig:m32-sim-x-effects-rmse-full}
\end{figure}

\begin{figure}[!htb]
    \centering
    \includegraphics[width = \linewidth]{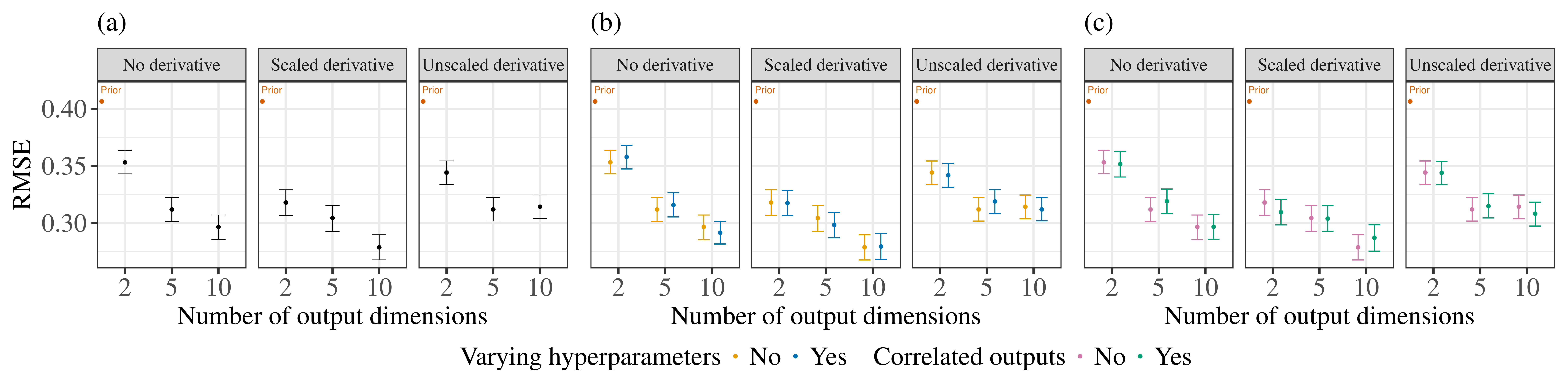}
    \caption{\textit{Matern 5/2 scenario: main effects of including (a) scaled derivatives and interaction effects of assuming (b) varying hyperparameters and (c) correlated outputs on recovery of latent inputs (full version) using RMSE.}}
    \label{fig:m52-sim-x-effects-rmse-full}
\end{figure}

\begin{figure}[!htb]
    \centering
    \includegraphics[width = \linewidth]{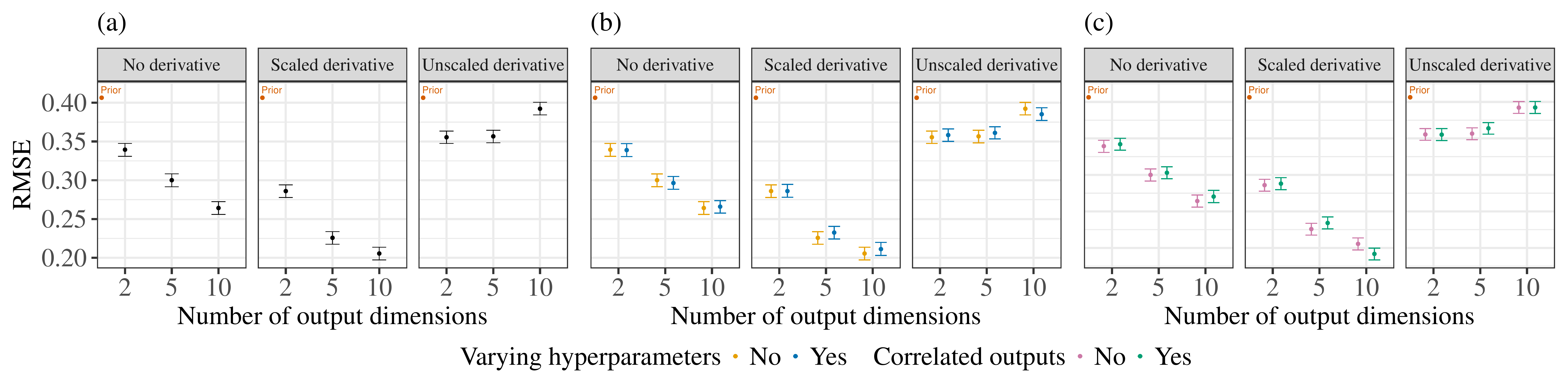}
    \caption{\textit{Periodic scenario: main effects of including (a) scaled derivatives and interaction effects of assuming (b) varying hyperparameters and (c) correlated outputs on recovery of latent inputs (full version) using RMSE.}}
    \label{fig:per-sim-x-effects-rmse-full}
\end{figure}

\begin{figure}[!htb]
    \centering
    \includegraphics[width = \linewidth]{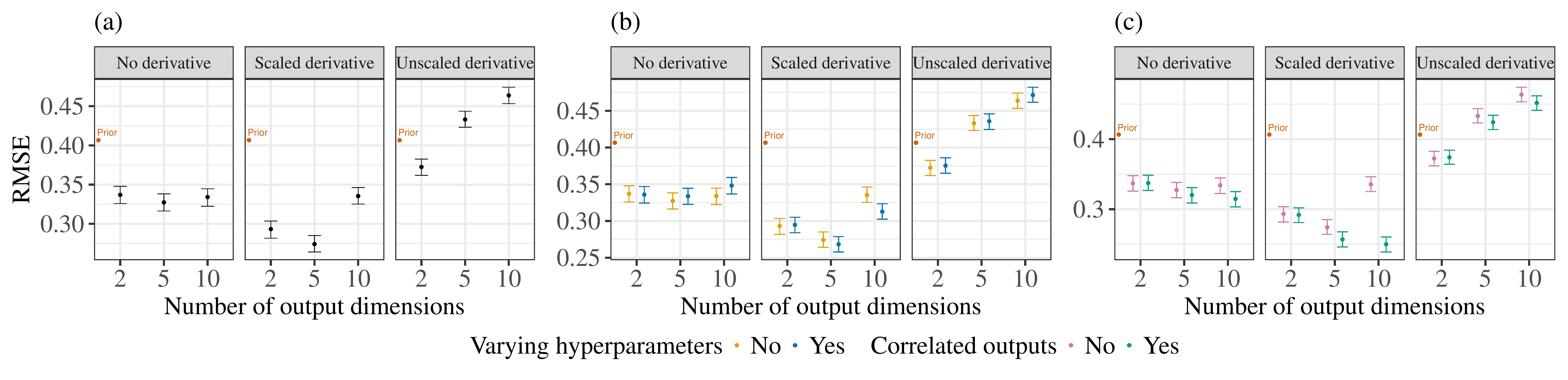}
    \caption{\textit{Periodic with trend scenario: main effects of including (a) scaled derivatives and interaction effects of assuming (b) varying hyperparameters and (c) correlated outputs on recovery of latent inputs (full version) using RMSE.}}
    \label{fig:per-trend-sim-x-effects-rmse-full}
\end{figure}
\newpage
\FloatBarrier
\subsubsection*{Full versions of model evaluation plots using MAE: Latent inputs} \label{app-latent-eval-mae}

\begin{figure}[!htb]
    \centering
    \includegraphics[width = \linewidth]{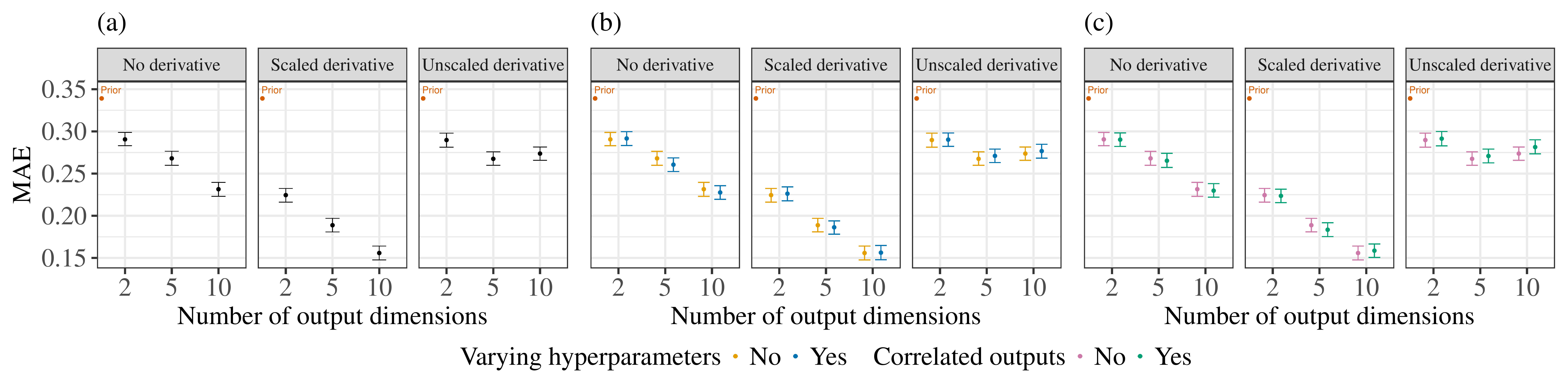}
    \caption{\textit{Squared exponential scenario: main effects of including (a) scaled derivatives and interaction effects of assuming (b) varying hyperparameters and (c) correlated outputs on recovery of latent inputs (full version) using MAE.}}
    \label{fig:se-sim-x-effects-mae-full}
\end{figure}

\begin{figure}[!htb]
    \centering
    \includegraphics[width = \linewidth]{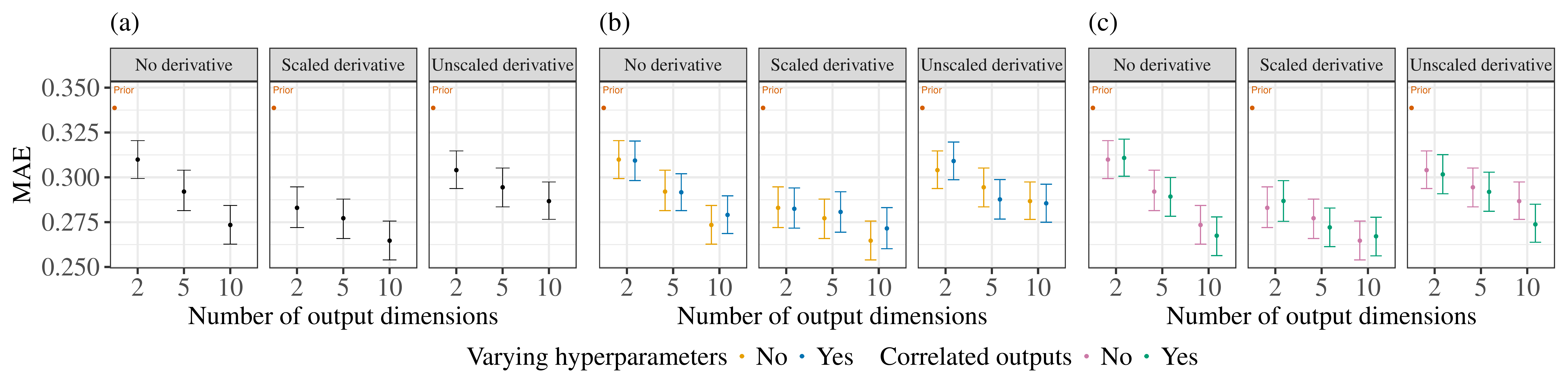}
    \caption{\textit{Matern 3/2 scenario: main effects of including (a) scaled derivatives and interaction effects of assuming (b) varying hyperparameters and (c) correlated outputs on recovery of latent inputs (full version) using MAE.}}
    \label{fig:m32-sim-x-effects-mae-full}
\end{figure}

\begin{figure}[!htb]
    \centering
    \includegraphics[width = \linewidth]{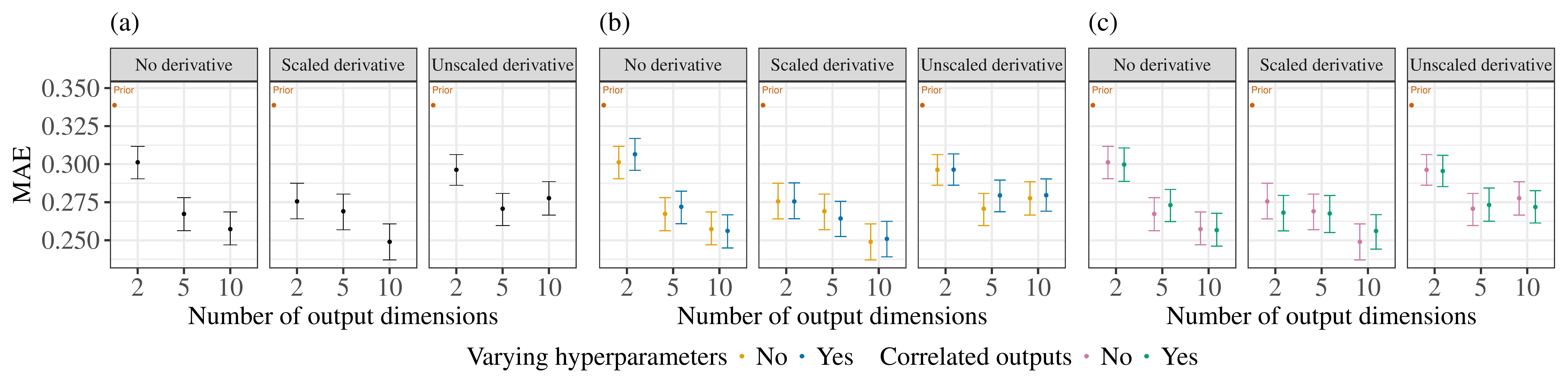}
    \caption{\textit{Matern 5/2 scenario: main effects of including (a) scaled derivatives and interaction effects of assuming (b) varying hyperparameters and (c) correlated outputs on recovery of latent inputs (full version) using MAE.}}
    \label{fig:m52-sim-x-effects-mae-full}
\end{figure}

\begin{figure}[!htb]
    \centering
    \includegraphics[width = \linewidth]{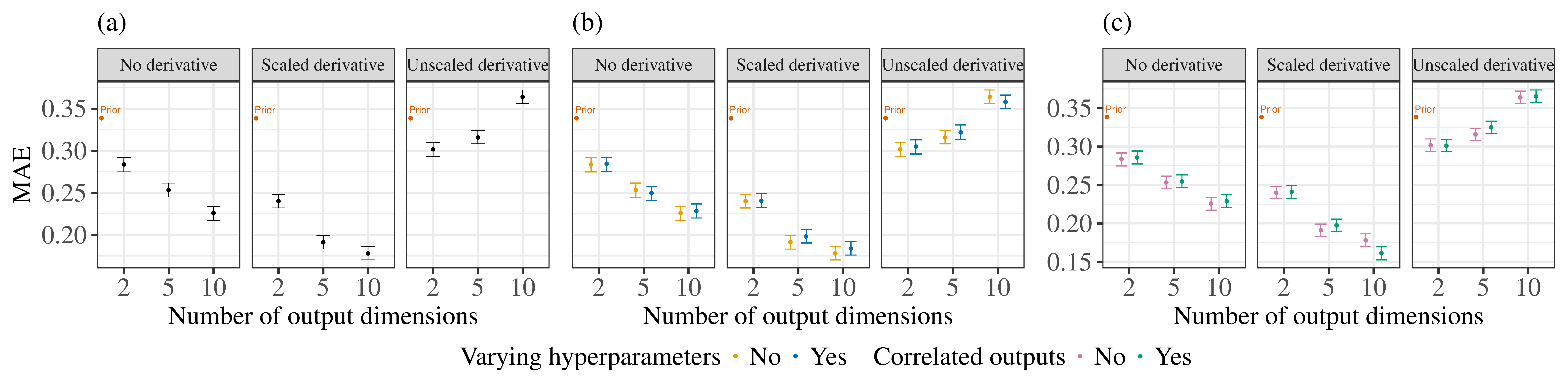}
    \caption{\textit{Periodic scenario: main effects of including (a) scaled derivatives and interaction effects of assuming (b) varying hyperparameters and (c) correlated outputs on recovery of latent inputs (full version) using MAE.}}
    \label{fig:per-sim-x-effects-mae-full}
\end{figure}

\begin{figure}[!htb]
    \centering
    \includegraphics[width = \linewidth]{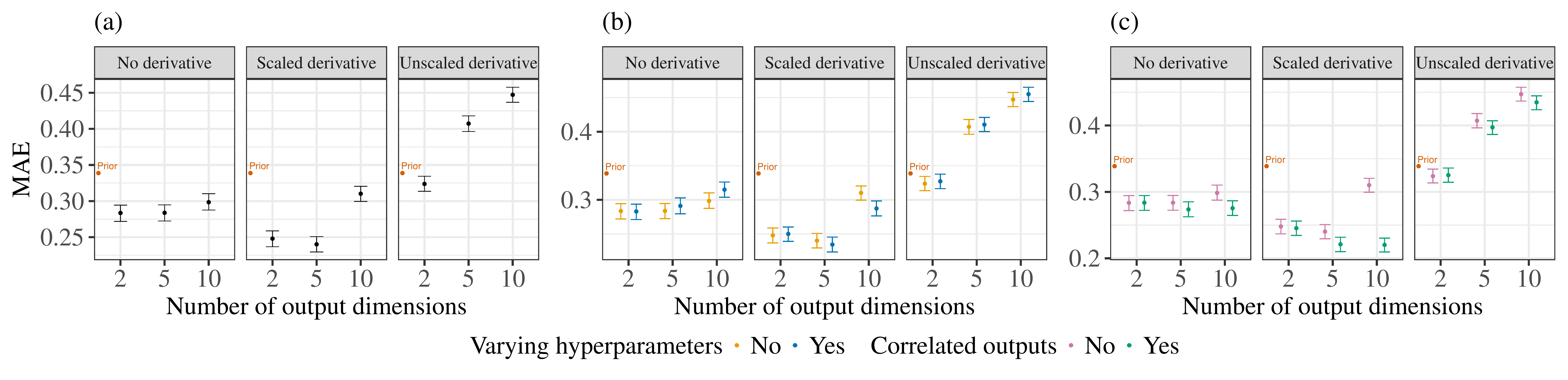}
    \caption{\textit{Periodic with trend scenario: main effects of including (a) scaled derivatives and interaction effects of assuming (b) varying hyperparameters and (c) correlated outputs on recovery of latent inputs (full version) using MAE.}}
    \label{fig:per-trend-sim-x-effects-mae-full}
\end{figure}

\newpage
\FloatBarrier
\subsubsection*{Additional model evaluation plots using RMSE:hyperparameters} \label{app-hyperparams-eval-rmse}

\begin{figure}[!ht]
    \centering
    \includegraphics[width = \linewidth]{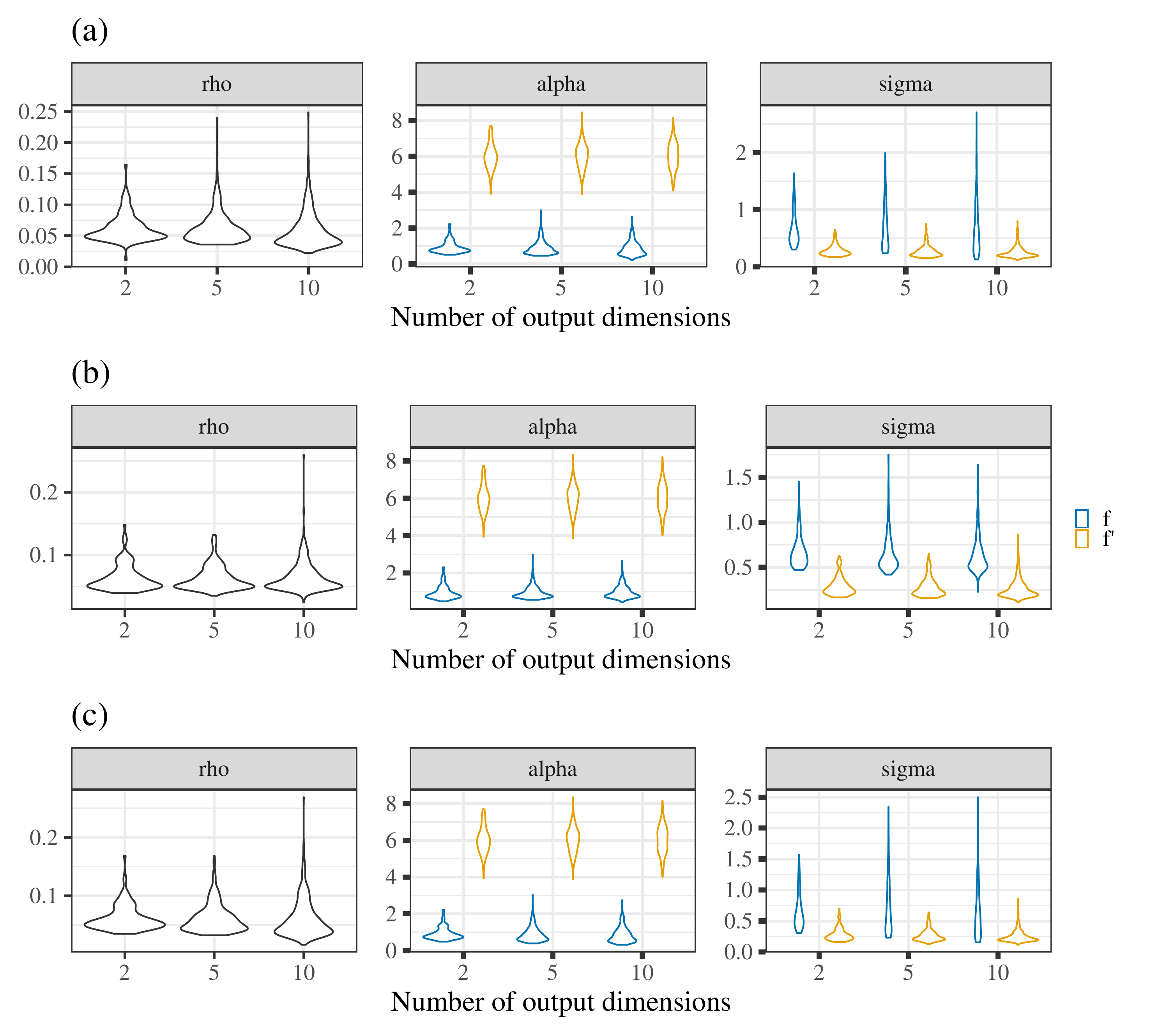}
    \caption{\textit{Squared exponential scenario: Hyperparameter RMSEs for scaled derivatives (a) without varying hyperparameters, (b) without correlated outputs and (c) without both varying hyperparameters and correlated outputs. The different colour denotes if the hyperparameters correspond to the original or the derivative part of the model.}}
    \label{fig:app-se-hyperparams-others}
\end{figure}

\begin{figure}[!ht]
    \centering
    \includegraphics[width = \linewidth]{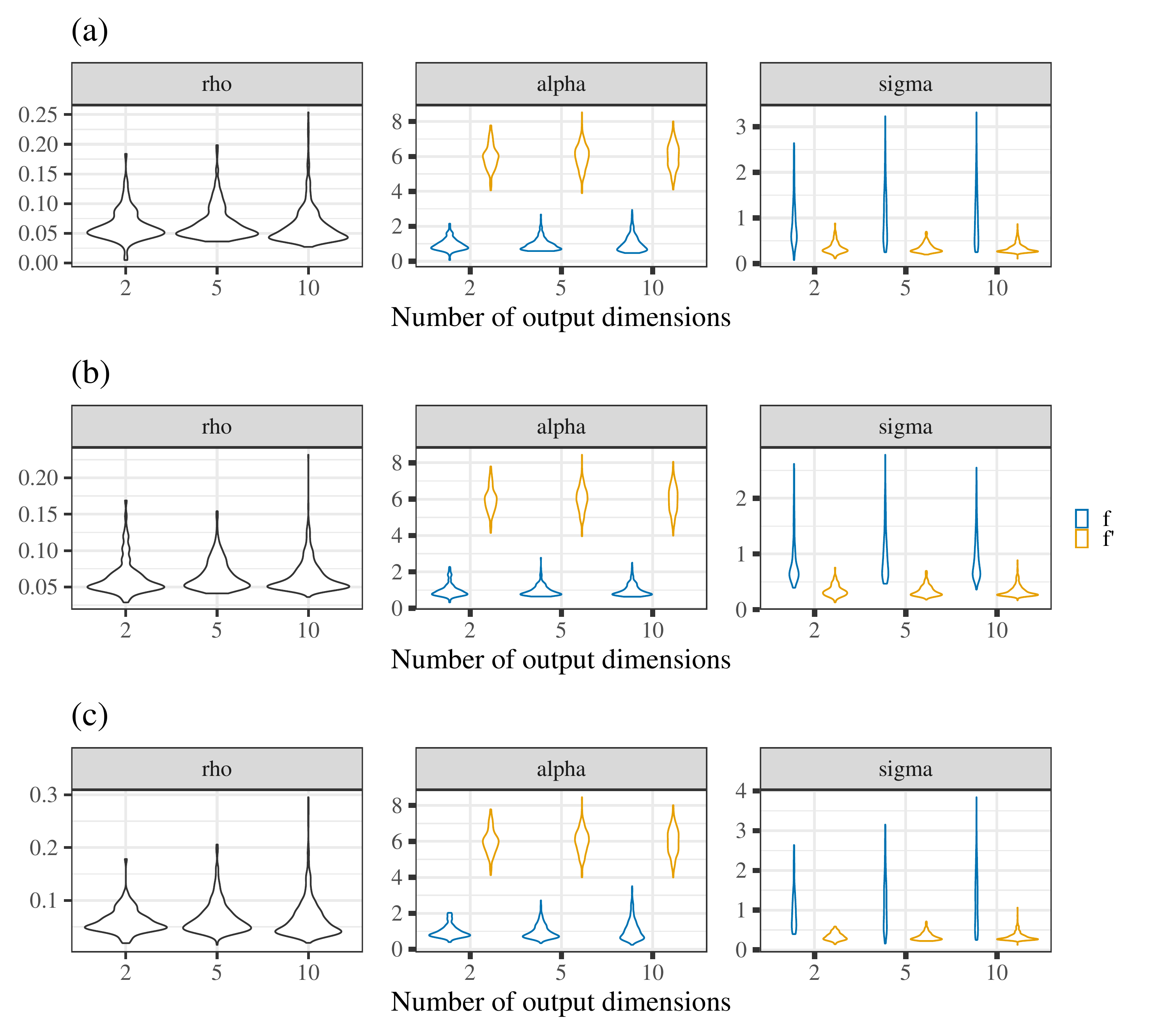}
    \caption{\textit{Matern 3/2 scenario: Hyperparameter RMSEs for scaled derivatives (a) without varying hyperparameters, (b) without correlated outputs and (c) without both varying hyperparameters and correlated outputs. The different colour denotes if the hyperparameters correspond to the original or the derivative part of the model.}}
    \label{fig:app-m32-hyperparams-others}
\end{figure}

\begin{figure}[!ht]
    \centering
    \includegraphics[width = \linewidth]{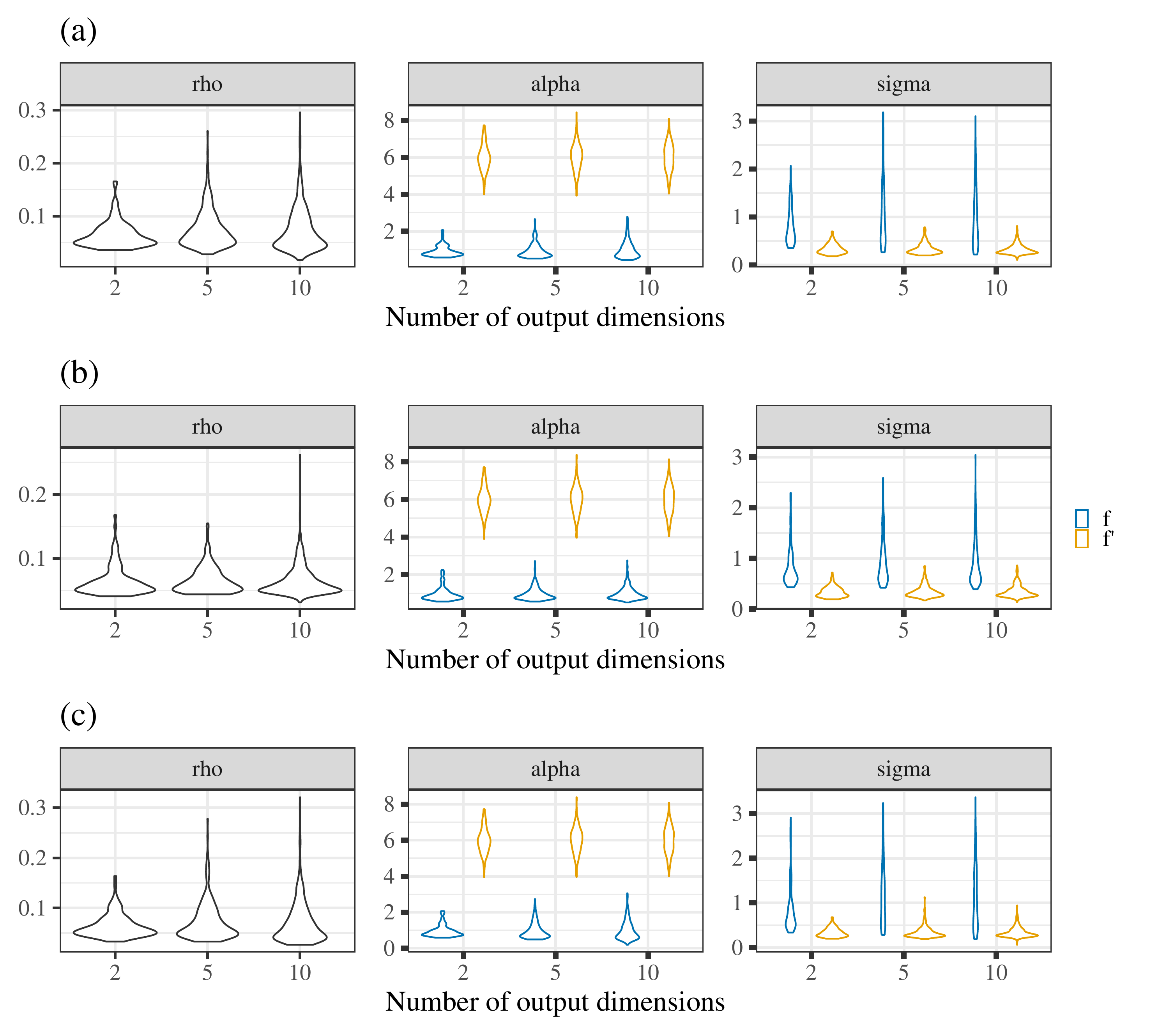}
    \caption{\textit{Matern 5/2 scenario: Hyperparameter RMSEs for scaled derivatives (a) without varying hyperparameters, (b) without correlated outputs and (c) without both varying hyperparameters and correlated outputs. The different colour denotes if the hyperparameters correspond to the original or the derivative part of the model.}}
    \label{fig:app-m52-hyperparams-others}
\end{figure}

\begin{figure}[!ht]
    \centering
    \includegraphics[width = \linewidth]{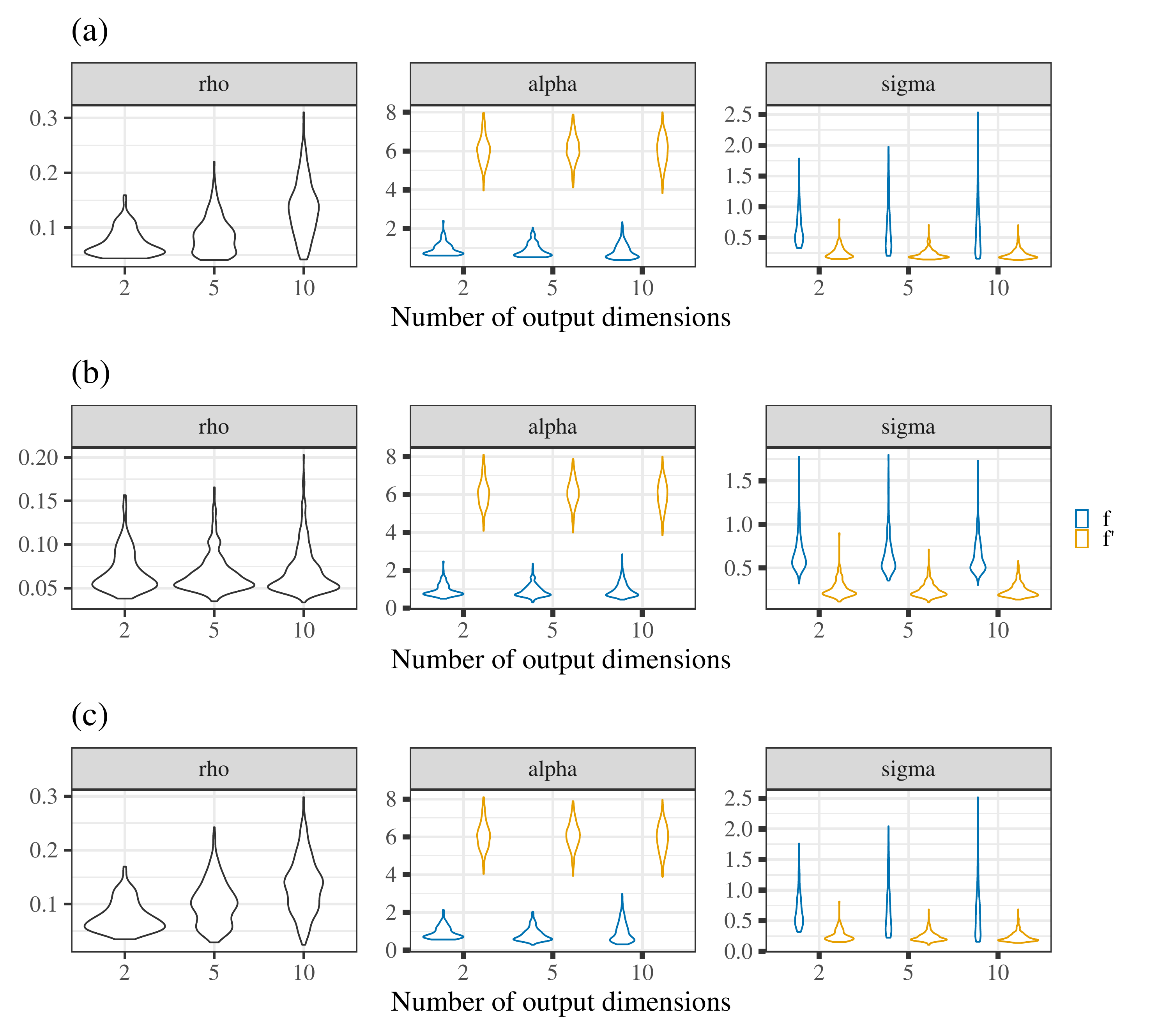}
    \caption{\textit{Periodic scenario: Hyperparameter RMSEs for scaled derivatives (a) without varying hyperparameters, (b) without correlated outputs and (c) without both varying hyperparameters and correlated outputs. The different colour denotes if the hyperparameters correspond to the original or the derivative part of the model.}}
    \label{fig:app-per-hyperparams-others}
\end{figure}

\newpage
\FloatBarrier
\subsubsection*{Model evaluation plot from simulation study with four MCMC chains} \label{app-mcmc-chains}

\begin{figure}[!ht]
    \centering
    \includegraphics[width=\linewidth]{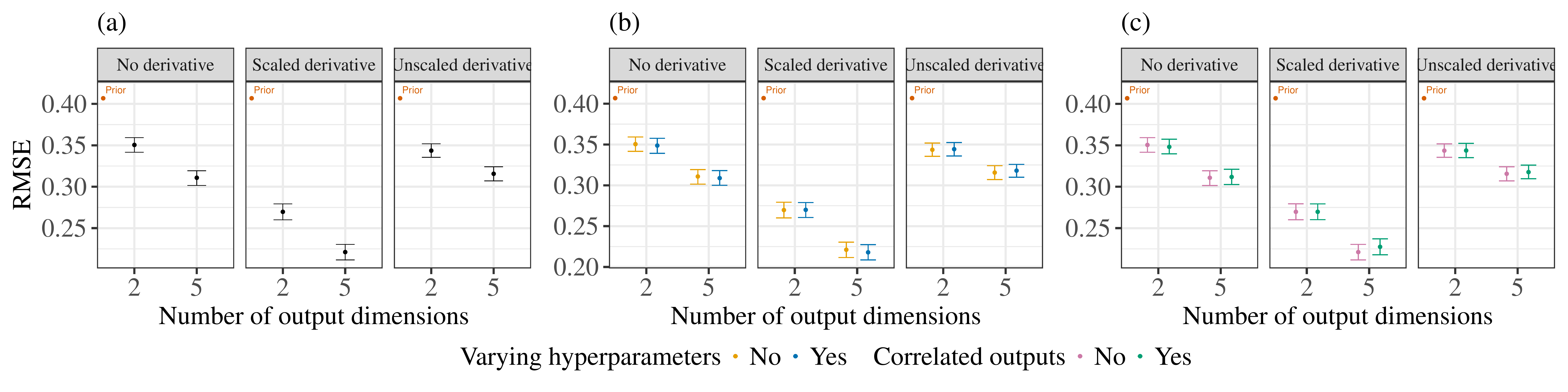}
    \caption{\textit{Squared exponential scenario with 4 MCMC chains: main effects of including (a) scaled derivatives and interaction effects of assuming (b) varying hyperparameters and (c) correlated outputs on recovery of latent inputs (full version) using RMSE.}}
    \label{fig:app-se-chains-x-effects-rmse-full} 
\end{figure}

\end{document}